\documentclass[twocolumn]{aastex631}
\usepackage{verbatim}
\usepackage{hyperref}
\hypersetup{
    colorlinks=true,
    linkcolor=blue,
    filecolor=magenta,      
    urlcolor=cyan,
    citecolor=blue,
}
\usepackage{amsmath}
\usepackage{comment}
\usepackage{graphicx}
\usepackage{tabularray}
\usepackage{xcolor}

\newcommand{\C}{\ensuremath{^{\circ}C\;}}

\newcommand{\ecosw}{\ensuremath{\sqrt{e}\cos\omega}}

\newcommand{\esinw}{\ensuremath{\sqrt{e}\sin\omega}}

\newcommand{\mstar}{\ensuremath{M_\star}}

\newcommand{\rearth}{\ensuremath{R_\earth}}

\newcommand{\rpl}{\ensuremath{R_{p}}}
\newcommand{\rstar}{\ensuremath{R_\star}}

\shorttitle{Catalog of TESS TTVs}
\shortauthors{Nabbie et al.}

\begin{document}

\title{Transit Timing Variations in \textit{TESS}: A Catalog from the First Five Years}

\author[0000-0003-0571-2245]{Emma Nabbie}
\affiliation{University of Southern Queensland, West St, Darling Heights, Toowoomba, Queensland, 4350, Australia}
\author[0000-0003-0918-7484]{Chelsea~X.~Huang}
\affiliation{University of Southern Queensland, West St, Darling Heights, Toowoomba, Queensland, 4350, Australia}
\author[0000-0001-9957-9304]{Robert A. Wittenmyer}
\affiliation{University of Southern Queensland, West St, Darling Heights, Toowoomba, Queensland, 4350, Australia}
\author[0000-0002-4891-3517]{George Zhou}
\affiliation{University of Southern Queensland, West St, Darling Heights, Toowoomba, Queensland, 4350, Australia}

\begin{abstract}

We present the first catalog of transit timing variations (TTVs) in \textit{TESS} systems with multiple \textit{TESS} Objects of Interest (TOIs) using data from Sector 1 to Sector 69, spanning the first five years of mission operations. With an initial sample of 175 multi-TOI systems, we find significant TTVs in 20 systems, 13 of which had not been previously detected. Our results are generally consistent with the findings of previous \textit{Kepler} TTV catalogs, with compact systems more likely to have detectable TTVs. However, the TTV systems in \textit{TESS} exhibit a pile-up at the 2:1 orbital period resonance, in contrast to the pile-up near the 3:2 resonance from previous \textit{Kepler} catalogs. This provides a tentative indicator that there may be different disk migration recipes that \textit{Kepler} systems favor versus \textit{TESS} systems. This catalog is a vital first step in determining which orbital resonances migrating planets tend to occupy at the end of formation, and aims to provide a list of high-impact targets for future in-depth follow up.

\end{abstract}

\keywords{planetary systems, planets and satellites: detection, techniques: photometric}

\section{Introduction}

NASA's \textit{Transiting Exoplanet Survey Satellite} \citep[\textit{TESS};][]{ricker} has surveyed nearly 95\% of the sky since its launch in 2018, with over 7000 \textit{TESS} Objects of Interest (TOIs) discovered. Of these, several systems have been found to exhibit transit timing variations (TTVs; e.g. TOI-216 \citep{Kipping:2019, Dawson:2021}, TOI-2525 \citep{Trifonov:2023}, TOI-1130 \citep{Huang:2020b}, TOI-4504 \citep{Vitkova:2025}). TTVs are due to gravitational interactions between planets, and the strength of this effect scales with proximity to a first-order mean-motion resonance (MMR). TTVs are valuable in both characterizing a system and tracing its origins, as resonant orbits are a signature of planetary migration through the gas disk \citep{Malhotra:1993,MustillWyatt:2011,Pichierri:2018}.

Previous systematic catalogs of TTVs in \textit{Kepler} have found hundreds of systems with aperiodic orbits \citep{Mazeh:2013,Rowe:2015,Holczer:2016}, with a pile-up at the 3:2 resonance. It is especially interesting to investigate if this pile-up is dynamically-driven, as resonance can reveal the relative parameters that govern how disk migration proceeds \citep{TerquemPapaloizou:2007, MustillWyatt:2011, OgiharaKobayashi:2013, Petrovich:2014, WangJi:2017}. For instance, 3:2 resonances are favorably-formed at later stages of formation, through higher migration speeds and slower accretion rates; 2:1 resonances are instead favored to form at earlier stages and with slower-migrating planets \citep{WangJi:2014}. However, it is not enough to simply study planets that are coincidentally near resonance. Instead, the presence of TTVs will confirm that a system's observed architecture is a consequence of migration.

\textit{TESS} provides a more diverse planetary sample than those found by \textit{Kepler}, as \textit{TESS} probes a broader stellar population with generally brighter stars. Therefore, a systematic search of \textit{TESS} TTVs is especially valuable to supplement previous \textit{Kepler} catalogs. Using \textit{TESS}, we examine the TTVs of a more general planet population to determine the fraction of planets that occupy different resonances. This is a crucial first step in investigating the dominant disk migration recipes by observing the resonances that planets tend to reside in by the end of migration.

A previous catalog of \textit{TESS} TTVs was conducted by \citet{Naponiello:2025}, however this was limited to systems with a single, confirmed transiting planet. In contrast, our sample includes all \textit{TESS} systems with multiple transiting planets, confirmed or candidate. This complements the previous systematic analysis by probing more compact architectures, while \citet{Naponiello:2025} places focus upon the search for non-transiting companions.

We present here the first catalog of \textit{TESS} TTVs in solely multi-TOI systems. Section 2 discusses the \textit{TESS} data and methodology used to construct our target list. Section 3 details the TTV search pipeline, including our procedure for light curve detrending and TTV fitting. Section 4 lists systems with significant TTVs, both new and previously-published, and discusses the significance of the catalog and future work.

\section{\textit{TESS} Data}

Our initial population was generated from the list of all TOIs as of 4 March, 2024, taken from the \textit{TESS} Exoplanet Follow-Up Observing Program (ExoFOP) website\footnote{\url{https://exofop.ipac.caltech.edu/tess/view_toi.php}}. In this list, we exclude any TOIs that were designated as false alarms, false positives, eclipsing binaries, or ambiguous planetary candidates. This is to help ensure that our sample only includes genuine transiting planets. We note that the Mikulski Archive for Space Telescopes (MAST) released reprocessed data from Sectors 48-65 in 2025, which aimed to increase photometric precision, but we do not use the reprocessed data in this work. This is because we do not report marginal detections, as our focus is on planets with the highest per-transit SNR, which would not be significantly affected by the increased photometric precision.

Moreover, we only consider light curves from the first five years of operations (Sectors 1-69). These include Full Frame Images (FFIs), which have varying observing cadences: 30 minutes for Cycles 1 and 2 (Sectors 1-26), 10 minutes for Cycles 3 and 4 (Sectors 27-55), and 200 seconds for Cycle 5 and onward (Sector 56+). Additionally, the \textit{TESS} Science Processing Operations Center \citep[SPOC;][]{Jenkins:2016} provides 2-minute light curves for selected targets. Of the FFIs and SPOC light curves, we use the shortest-cadence data available per sector for each target star that we examine.

\subsection{TOI Selection Criteria}
\label{sec:criteria}

To construct the catalog, we place several limits on the initial population of known TOIs to exclude systems where TTV studies would be infeasible. Our selection criteria are as follows:

\begin{enumerate}
    \item We implement a host star magnitude cut for \textit{TESS} magnitude $T_\text{mag} < 12$ for sufficient photometric precision to detect transiting planets.
    \item Each system must host multiple TOIs as of Sector 69. Single-TOI systems are not considered within this catalog, as searching for the TTV signals due to non-transiting companions is beyond the scope of this work.
    \item Planets in selected systems must have at least four transits in Sectors 1-69 for TTV analysis. This is to mitigate the possibility of detecting spurious TTVs from imprecise ephemerides.
\end{enumerate}

Our final sample consists of 175 multi-TOI systems that satisfy all of the above criteria. Of these, 20 systems host at least one giant planet (which we define here as $R > 8.5 R_{\oplus}$), while the rest only host smaller planets.

\section{TTV search pipeline}

\subsection{Light Curve Processing}

The first step in our pipeline is to access each sector of \textit{TESS} data for a given TOI using the {\tt lightkurve} package's {\tt search\_lightcurve()} function \citep{lightkurve, astroquery}. When available, we use shorter-cadence SPOC light curves over those generated from FFIs by the \textit{TESS} Quick-Look Pipeline \citep[QLP;][]{Huang:2020b}. For SPOC data, we use the Pre-Search Data Conditioned Simple Aperture Photometry (PDCSAP) flux using the {\tt PDCSAP\_FLUX} header, which accounts for instrument systematics like dilution or blending. We then select the data with the shortest exposure time, selecting a quality mask that removes all points with quality flags.

After normalizing the Simple Aperture Photometry (SAP) flux, we perform mean absolute deviation (MAD) clipping for above-transit outliers. We exclude all points that are five MADs above the median flux value. This removes any observed flares or systematic effects. Other artifacts, such as spacecraft momentum dumps, had to be manually removed on a case-by-case basis. Our detrending is performed with {\tt keplersplinev2} \footnote{\url{https://github.com/avanderburg/keplersplinev2}} \citep{VanderburgJohnson:2014, ShallueVanderburg:2018} on the out of transit portion of the light curves, cutting out 0.75 transit durations on each side of the predicted transit centers of all planets. This is to prevent transits from being modeled out of the light curve during detrending. 

This initial iteration of detrending is used to generate additional masks to remove any points more than 6{$\sigma$} from the median flux value on the pre-detrended light curve. This is to remove any outlier points that may bias detrending. A second \textit{Kepler} spline is performed on the original light curve including these new masks. This spline is then interpolated to include the in-transit points, and we apply the interpolated spline to the original raw, normalized light curve. We perform this detrending algorithm on each sector of data separately, then combine all sectors to produce the light curves that are used as inputs for our TTV search.

\subsection{Initializing MCMC}

We initialize our Markov Chain Monte Carlo (MCMC) routine with the following free planet parameters: period $p$, {${\rpl}/{\rstar}$}, impact parameter $b$, {\esinw}, {\ecosw}, and transit center $t_0$ at each epoch. Stellar parameters {\mstar}, {\rstar}, and limb darkening parameters $q_1$ and $q_2$ were also left free. 

\begin{figure*}
    \centering
    \includegraphics[width=0.8\linewidth]{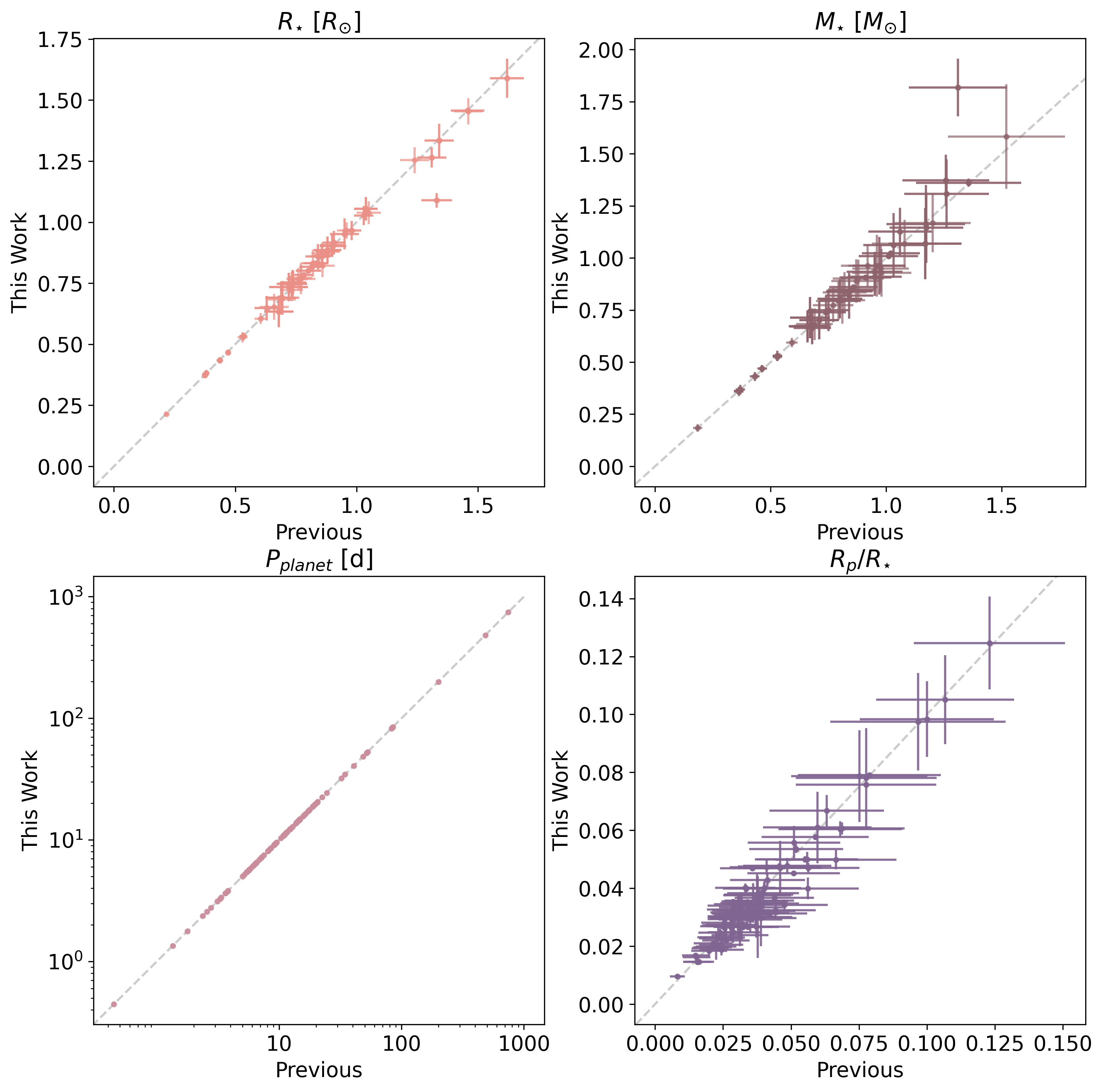}
    \caption{Correlation plots showing the stellar radius (top left), stellar mass (top right), orbital period (bottom left), and planet-star radius ratio (bottom right) derived in this work versus in the TOI catalog. Values from previous published literature were used instead of TIC values where available. One-sigma errorbars are denoted by horizontal and vertical bars, with a dashed $x=y$ line to guide the eye.}
    \label{fig:correlation}
\end{figure*}

All stellar and planet parameter prior values were taken from \textit{TESS} Input Catalog \citep[TIC v8.2;][]{Stassun:2019,Paegert:2021}. A correlation plot between the stellar and planet parameters derived in this work versus those from TIC (or previous publications, where relevant) are shown in Figure \ref{fig:correlation}. We constrain the stellar mass and radius with a gaussian prior, initialized at the TIC value and error for each parameter. If the errors for stellar mass and radius are not available, we implement an error of 0.1 $M_\odot$ and $R_\odot$, respectively. We fit for limb darkening parameters {$q_{1}$} and {$q_{2}$} and recovered limb darkening coefficients {$u_{1}$} and {$u_{2}$} through equations 15 and 16 of \cite{Kipping:2013}. Both {$q_{1}$} and {$q_{2}$} were constrained with uniform priors between 0 and 1.

Priors on planet orbital periods were also constrained with a gaussian prior, defined by the values in the TOI catalog. An error of 0.1 days was also used if the error on the period was not supplied by TOI catalog. The planet-to-star radius, {\rpl}/{\rstar}, was bound with a uniform prior between 0 and 1. If a planet's radius is not available within the TOI catalog, the planet is skipped. The impact parameters $b_i$ for both planets are bounded between 0 and 1. Instead of fitting for the planets' eccentricities $e_i$ and longitudes of periastron $\omega_i$, we fit for the parameters {\esinw} and {\ecosw} to account for the degeneracy between $e$ and $\omega$. These parameters were uniformly bounded between -1 and 1. 

To generate transit time priors, we begin with predictions using linearly-propagated transit centers. At each observed epoch, we cut out a segment of the light curve with 10 transit durations on each side, then perform a least squares fit to constrain the transit center $t_0$. The best-fit transit time from the least squares fit is then used to generate a uniform box prior on the transit times, with 10$\sigma$ bounds on each side, where $\sigma$ is taken from the TOI catalog for its ephemeris. It should be noted that we only fit for epochs in which 80\% of the transit has been observed so that the transit center can be sufficiently determined.

For targets with a per-transit signal-to-noise ratio (SNR) of less than 5, we assign a single transit center to an entire sector of data. Therefore, we perform the same process as outlined above, except we fit each sector separately rather than each transit epoch separately. Each sector has a linearly-predicted transit time, which we define from the TIC $t_0$ value and implement a generous box prior with a width of 0.1 days. This loose prior avoids over-constraining the per-sector transit centers during our MCMC fit. 

For the systems that we know to have transit events outside of the predicted 0.1-day window due to extremely large TTVs (via visual inspection or previous literature), we manually broaden the prior window for the transit centers. This special handling was only necessary in less than 5 systems, and all target planets were higher-SNR. We also use published ephemerides as priors wherever possible for these particular systems.

\subsection{Deriving TTVs}

After defining priors and processing the light curves, we classify planetary systems based on the per-transit SNR of member planets. The first category considers ``High-SNR" systems, where all planets have a per-transit SNR $>$ 5. ``Medium-SNR" systems have at least one planet on either side of the SNR $< 5$ cut-off. In ``Low-SNR" systems, all member planets have a per-transit SNR $<$ 5. We define the SNR cut-off value based on the threshold defined in \citet{Ioannidis:2016}, and scale our value to reflect the per-transit SNR rather than the overall transit SNR. Below our defined value, we cannot perform per-transit TTV fitting, as the individual transits do not have sufficient SNR to constrain the transit center.

We use the {\tt emcee} package \citep{Foreman-Mackey2013} to model all the transit signals. We obtain best-fit planet parameters and transit times by fitting a {\tt batman} \citep{Kreidberg(2015)} transit model to \textit{TESS} data. For High-SNR systems, we fit each transit epoch of each planet individually. Each transit model shares the same relevant planet and stellar parameters, save for the transit center. We also isolate double transit events, fitting multiple planet parameters and epochs jointly by using a combined {\tt batman} transit model. For Medium- and Low-SNR systems, we fit planets with SNR $<$ 5 on a per-sector basis. We treat an entire sector of data as a single transit epoch, and fit for planet parameters and transit times using the same method as described above. Therefore, due to the selection criteria in Section \ref{sec:criteria}, planets with SNR $<$ 5 must have at least four sectors of data to undergo fitting.

Moreover, we account for the varying cadences of \textit{TESS} observations by super-sampling the light curve using a factor of 7. Super-sampling divides each exposure into 7 even segments, producing the weighted mean flux for each segment so that the average integration time aligns with the observed value. This is to mitigate any change in the transit shape that comes from longer exposure times. 

\subsection{Identifying Statistically Significant TTVs}
\label{sec:significance_criteria}

We quantify significance of TTV signals based on the statistical deviation of individual transit times from a linear model. We also estimate the significance of TTVs using Lomb-Scargle periodograms \citep{Lomb:1976, Scargle:1982} for planets with more than 10 observed transits. However, this is not possible for the majority of systems. This is due to \textit{TESS} having a shorter baseline than that of \textit{Kepler}, leading to insufficient sampling of the TTV periodicity. 

We compare the peaks of the Lomb-Scargle periodograms to the 0.01\%, 0.1\%, and 0.5\% False Alarm Probabilities. These thresholds are calculated by testing the probability that a periodogram peak would be observed at a given power, assuming that the data had no periodic signals. We use the standard Lomb-Scargle algorithms from Astropy\footnote{\url{https://docs.astropy.org/en/latest/timeseries/lombscargle.html}} for these calculations \citep{astropy}. Planets with transit times that deviate from a linear ephemeris model by more than $5\sigma$, or whose periodograms peak above the 0.01\% False Alarm Probability (FAP), are classified as having statistically significant TTVs. We do not evaluate the significance of transit depth or duration variations, as this is beyond the scope of our initial catalog. 

\section{Results and Discussion}

\subsection{Systems with Significant TTVs}

Following the significance criteria laid out in Section \ref{sec:significance_criteria}, we find that of the 76 systems analyzed, 20 exhibit significant TTVs ($\sigma \geq 5$). Of these, 13 were not previously-known TTV systems. The individual and per-sector transit times for all significant-TTV systems can be found in the Appendix (Table \ref{tab:ttvfull}). Best-fit orbital parameters for all analyzed planets are summarized in Table \ref{tab:fullparams} of the Appendix. Machine-readable versions of these tables are also available.

\begin{figure*}[htp]
\centering
\includegraphics[width=.5\textwidth]{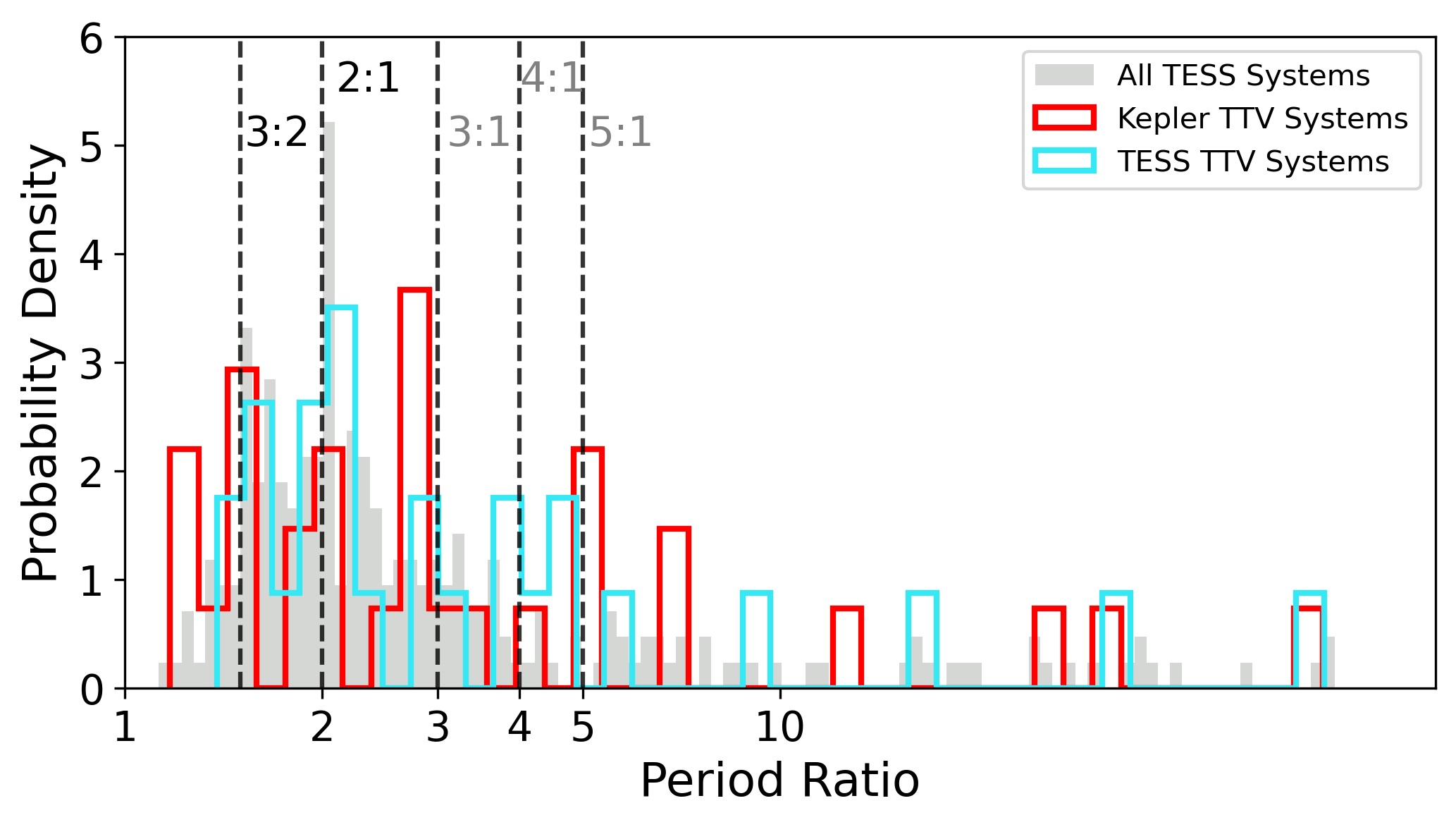}\hfill
\includegraphics[width=.5\textwidth]{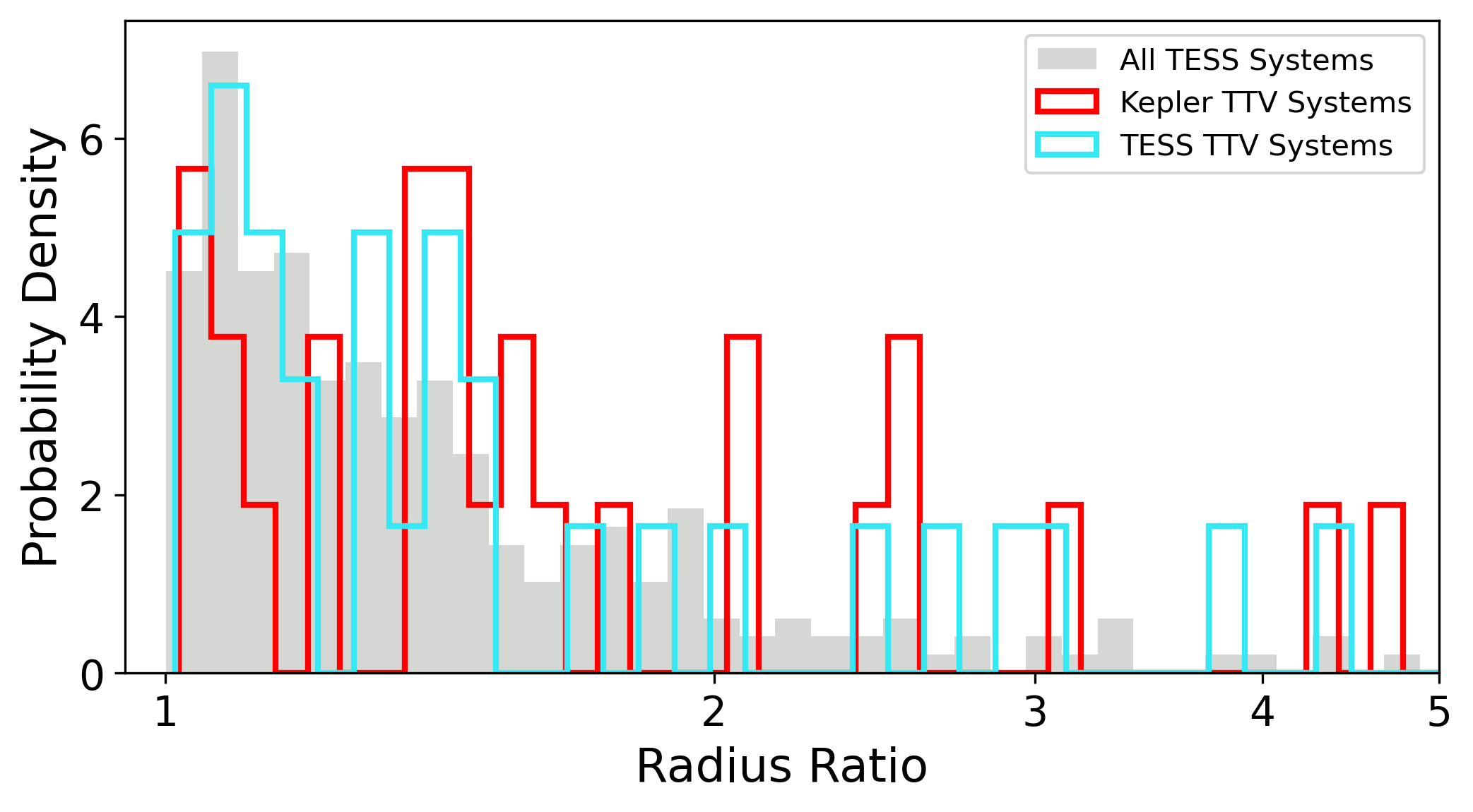}\hfill

\caption{\textit{Left:} Distributions of period ratios of neighboring planet pairs in \textit{TESS} and \textit{Kepler} systems. The general \textit{TESS} population is shown in grey, while \textit{TESS} TTV systems are shown in cyan. \textit{Kepler} TTV systems, taken from \citet{Holczer:2016}, are shown in red. The vertical dashed lines denote the locations of orbital resonances, with first-order resonances shown in black text. Higher-order resonances are in grey text. \textit{Right:} Distributions of neighboring planet radius ratios in \textit{Kepler} and \textit{TESS}, with the same color scheme.}
\label{fig:pratiohist}

\end{figure*}

\begin{figure*}[htp]
\centering
\includegraphics[width=.7\textwidth]{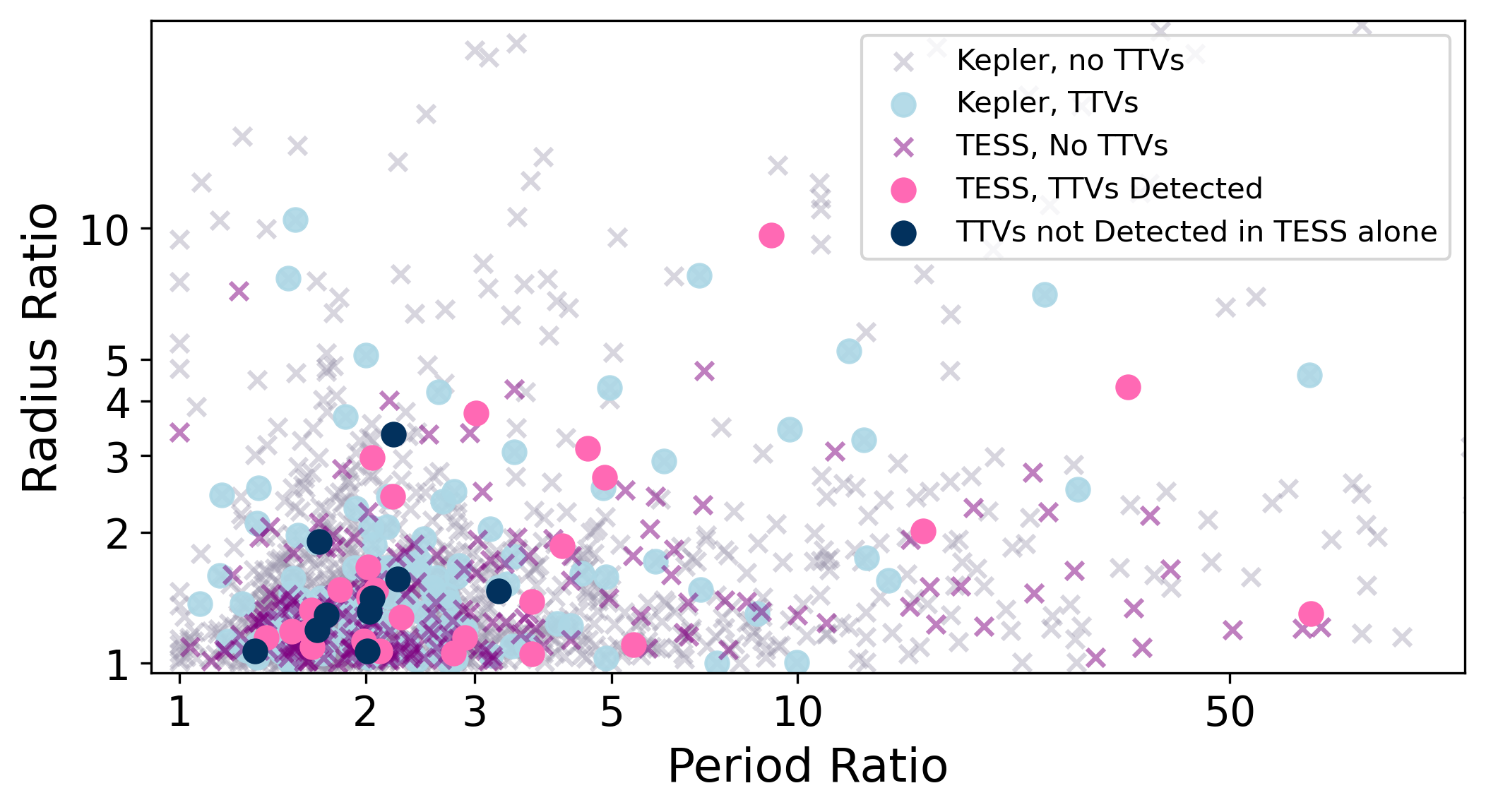}\hfill

\caption{The population of \textit{TESS} TTV systems in period ratio versus radius ratio space, in context with \textit{Kepler} TTV systems. \textit{TESS} and \textit{Kepler} systems without detectable TTVs are represented by x markers, while systems with detectable TTVs in \textit{TESS} and \textit{Kepler} are shown in pink and light blue, respectively. Systems with TTVs that are not detectable in \textit{TESS} alone are shown in dark blue.}
\label{fig:ttvscatter}

\end{figure*}

In general, we can rule out spot-induced TTVs in the systems we present, as we are not reporting marginal detections. Distributions of neighboring planet period and radius ratios are shown in Figure \ref{fig:pratiohist}, placing \textit{TESS} TTV systems in context with those from \textit{Kepler}. Figure \ref{fig:ttvscatter} depicts the period ratio vs radius ratio of \textit{TESS} and \textit{Kepler} systems, finding similar populations with TTVs between the two missions.

The Lomb-Scargle periodograms for the 10 planets with sufficient baseline are shown in Appendix \ref{appendix}, Figure \ref{fig:periodograms}. Of the 2 planets with sufficient baseline for Lomb-Scargle analysis, 3 did not pass the 0.01\% FAP threshold, but did show peaks in their periodograms above the 0.5\% FAP. These systems may have long-term TTVs, as the observed signals cannot be explained by any astrophysical false positive scenario. Future follow-up efforts with longer baselines will be crucial to measure the super-period of the observed TTVs.

\subsubsection{Known TTV Systems}

Our sample contains several previously-published TTV systems: 

\begin{itemize}
    \item \textbf{TOI-178:} This system hosts six planets, with TTVs predicted by the original discovery in \citet{Leleu:2021}, and then fully characterized in \citet{Leleu:2024} using additional data from CHEOPS and NGTS. With the available \textit{TESS} data, we are only able to analyze the TTVs for TOI-178\,d ($P = 6.56$ days, $R = 2.57R_\oplus$). Using 10 transits of TOI-178\,d, we find $\sim$30-minute TTVs with a predicted super-period of approximately 3000 days. This is consistent with the findings of the earlier TTV study, although the super-period in \citet{Leleu:2024} is found to be much shorter ($\sim260$ days). This is because the underlying sub-structure of the TTVs was only revealed using additional CHEOPS data, and is unseen in the TESS data.
    
    \item \textbf{TOI-201:} The TTVs of TOI-201\,b were characterized in \citet{Maciejewski:2025}, induced by perturbations from a 7.7-year giant transiting planet (TOI-201\,c). Our catalog finds 20-minute TTVs with TOI-201\,b that coincide with the transit of TOI-201\,c. This is different from the 30-minute amplitude found in \citet{Maciejewski:2025}, as that analysis used \textit{TESS} data up to Sector 88, whereas our catalog only uses data up to Sector 69. For the overlapping sectors of \textit{TESS} data, our results are nevertheless consistent.
    
    \item \textbf{TOI-216:} We find the high-amplitude TTVs for TOI-216\,b and c that were measured in previous analyses by \citet{Kipping:2019, Dawson:2021,  McKee:2023}. The TTV solution is consistent with previous literature, with only a handful of additional epochs added for TOI-216\,c.

    \item \textbf{TOI-1130:} Our search finds 2-hour TTVs for TOI-1130\,b and 10-minute TTVs for TOI-1130\,c. This matches the signals that were first reported in \citet{Huangtoi1130}. Our results are also consistent with those of subsequent TTV analyses from \citet{Korth:2023} and \citet{Borsato:2024}.

    \item \textbf{TOI-2525:} We find 6-hour and 50-minute TTVs for TOI-2525\,b and c, respectively. This matches the measured signals first detected in \citet{Trifonov:2023}.  

    \item \textbf{TOI-4504:} An in-depth TTV analysis was carried out by \citet{Vitkova:2025}, finding nearly 2-day amplitude TTVs for the warm Jupiter TOI-4504\,c. This led to the discovery of a third, non-transiting warm Jupiter in between the two planets observed by \textit{TESS} \citep{Vitkova:2025}. We also find $\sim$46-hour TTVs for TOI-4504\,c in our \textit{TESS} search, consistent with the extremely large TTVs previously measured.

    \item \textbf{HD 108236} This system hosts five planets, with TTVs evaluated in \citet{Hoyer:2022}; however, \textit{TESS} only provides sufficient SNR to evaluate the TTVs of HD 108236\,d ($P = 14.18$ days, $R = 2.54R_\oplus$) and HD 108236\,e ($P = 19.59$ days, $R = 3.08R_\oplus$). Using 4 transits each, we find $\sim$2 minute TTVs for HD 108236\,d, and $\sim$6 minute TTVs for HD 108236\,e. These signals do not exhibit coherent sinusoidal behavior, but instead scatter about a straight line at 6- and 23-$\sigma$ significance, respectively. This is consistent with the TTV analysis carried out in \citet{Hoyer:2022}.
\end{itemize}

\begin{figure*}[htp]
\centering
\includegraphics[width=.7\textwidth]{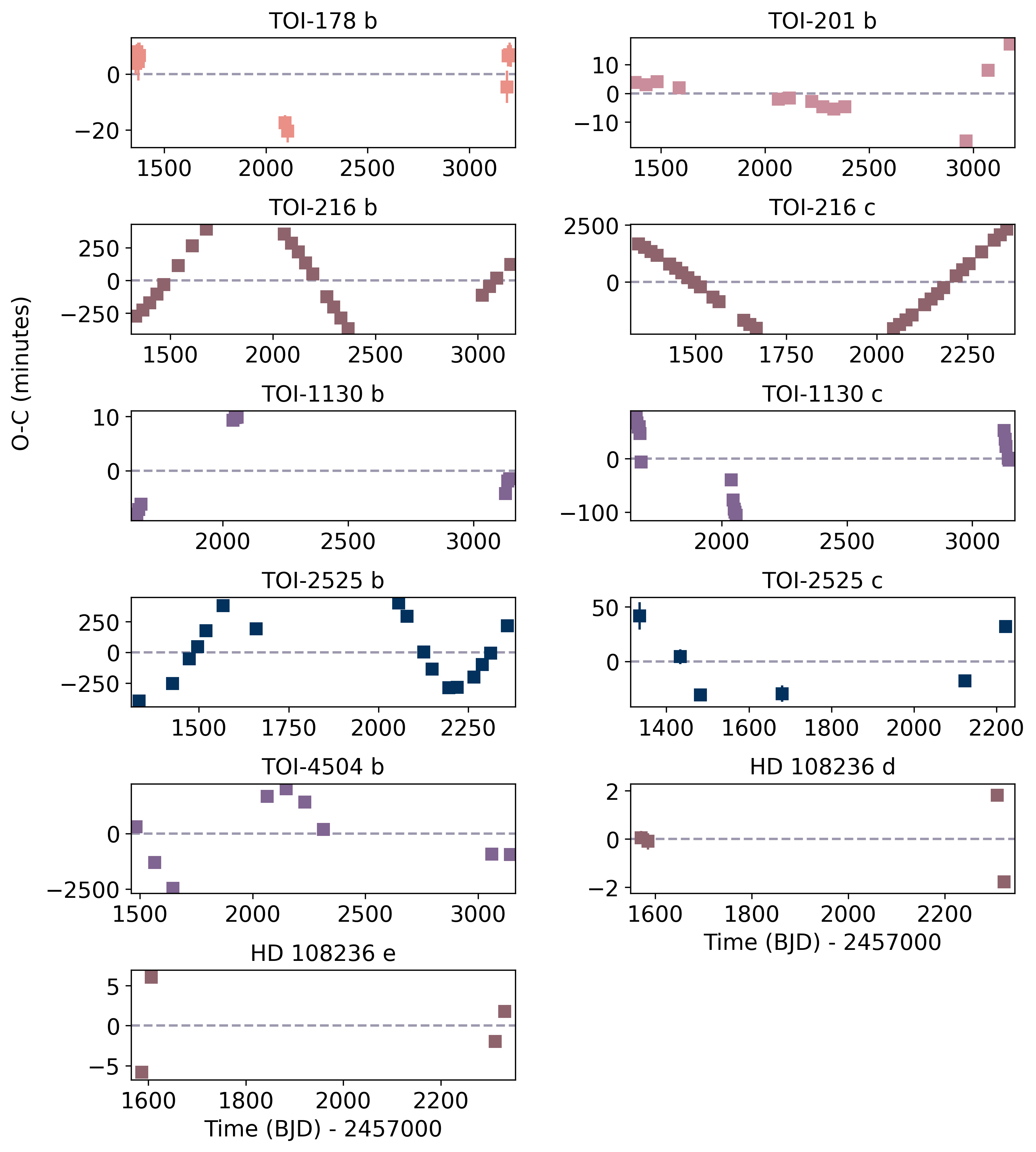}\hfill

\caption{Recovered signals of planets with previously-published TTVs. Matching colors indicate planets from the same system. Errorbars are included but too small to be visible in most cases. The above results are consistent with those from previous TTV analyses in the literature.}
\label{fig:known_ttvs}

\end{figure*}

Through our pipeline, we are able to recover all of the known TTV signals from these systems. We find consistent TTV solutions for systems with additional epochs of \textit{TESS} data (Table \ref{tab:publishedttvs}). Figure \ref{fig:known_ttvs} contains the measured TTV signals from all known TTV systems in our sample, which are consistent with the results of previous literature.

\begin{deluxetable*}{ccccccc}
\tablecaption{
  TTV Results in This Work Versus Previous Literature
  \label{tab:publishedttvs}
}
\tablehead{ 
    \colhead{Host Star} & \colhead{Planet} & \multicolumn{2}{c}{Approximate TTV Amplitude [min]} & \multicolumn{2}{c}{Estimated Super-Period [days]} & \colhead{Source}
}

\startdata
 & & \textbf{This work} & Published & \textbf{This work} & Published & \\ \hline
TOI-178 & d & 30 & 35 & 3000 & 260 & \citet{Leleu:2024}\\
TOI-201 & b & 20 & 30 & - & - & \citet{Maciejewski:2025}\\
TOI-216 & b & 540 & 540 & 1500 & 1500 & \citet{McKee:2023}\\
TOI-216 & c & 3200 & 3200 & 1500 & 1500 & \citet{McKee:2023}\\
TOI-1130 & b & 10 & 8 & - & - & \citet{Borsato:2024}\\
TOI-1130 & c & 120 & 120 & - & - & \citet{Borsato:2024}\\
TOI-2525 & b & 350 & 360 & - & - & \citet{Trifonov:2023}\\
TOI-2525 & c & 50 & 58 & - & - & \citet{Trifonov:2023}\\
TOI-4504 & b & 2880 & 2880 & 900 & 930 & \citet{Vitkova:2025}\\
HD 108236 & d & 2 & 2 & - & - & \citet{Hoyer:2022}\\
HD 108236 & e & 6 & 2 & - & - & \citet{Hoyer:2022}\\
\enddata
\end{deluxetable*}

\subsubsection{New TTV Systems}

We find 13 systems exhibiting significant TTVs that have not been previously detected in the literature. We measure and report the TTV signals here, however we note that it is beyond the scope of this preliminary catalog to rule out all astrophysical false positive scenarios for each individual system. Further follow up will be needed to confirm the nature of these signals. 

\begin{figure*}[htp]
\centering
\includegraphics[width=.7\textwidth]{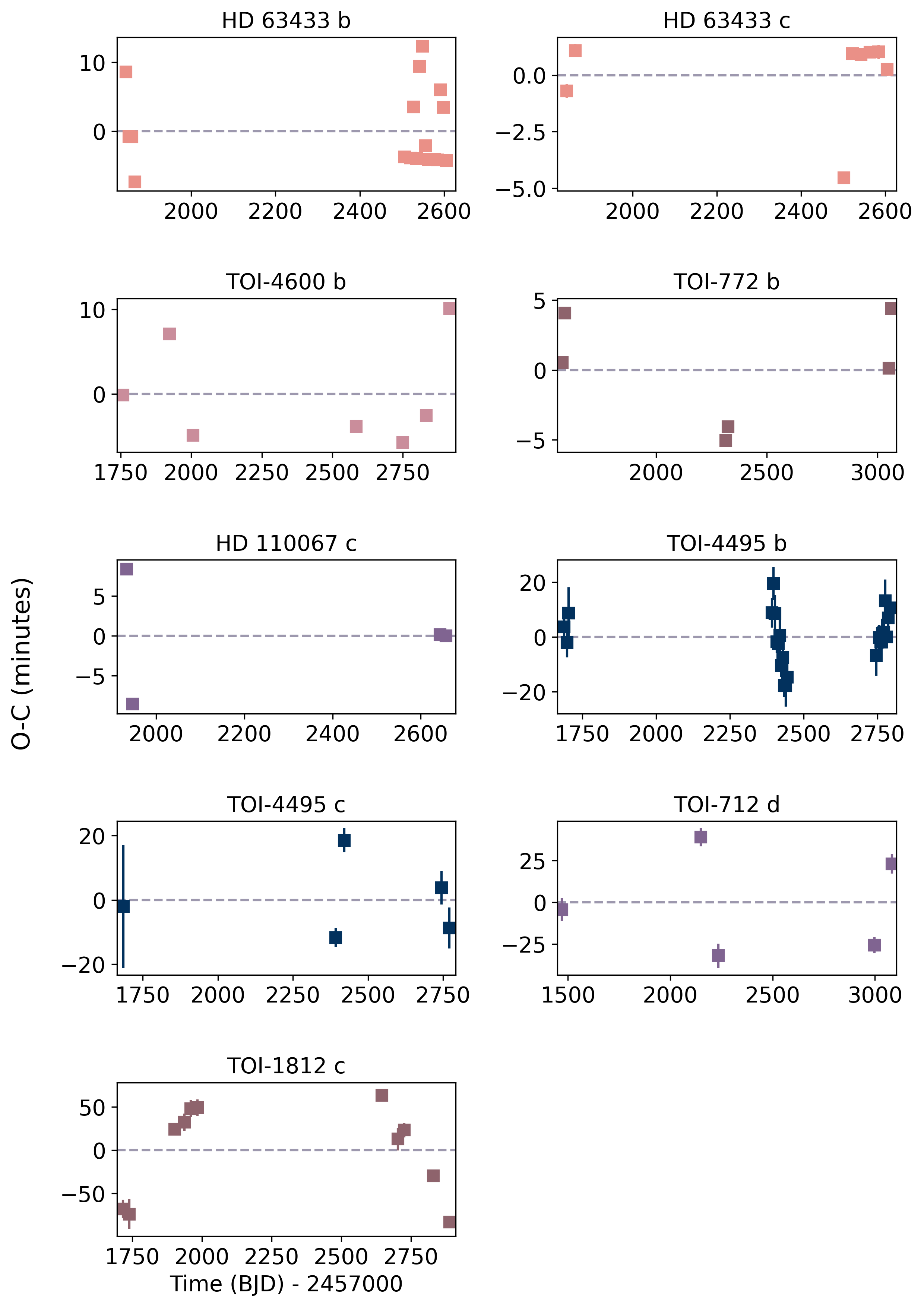}\hfill

\caption{TOIs with significant TTVs in host systems where TTVs had not been previously measured. One-sigma errorbars are included but may be too small to be visible. The color scheme is shared with Figure \ref{fig:known_ttvs}.}
\label{fig:new_ttvs1}

\end{figure*}

\begin{figure*}[htp]
\centering
\includegraphics[width=.7\textwidth]{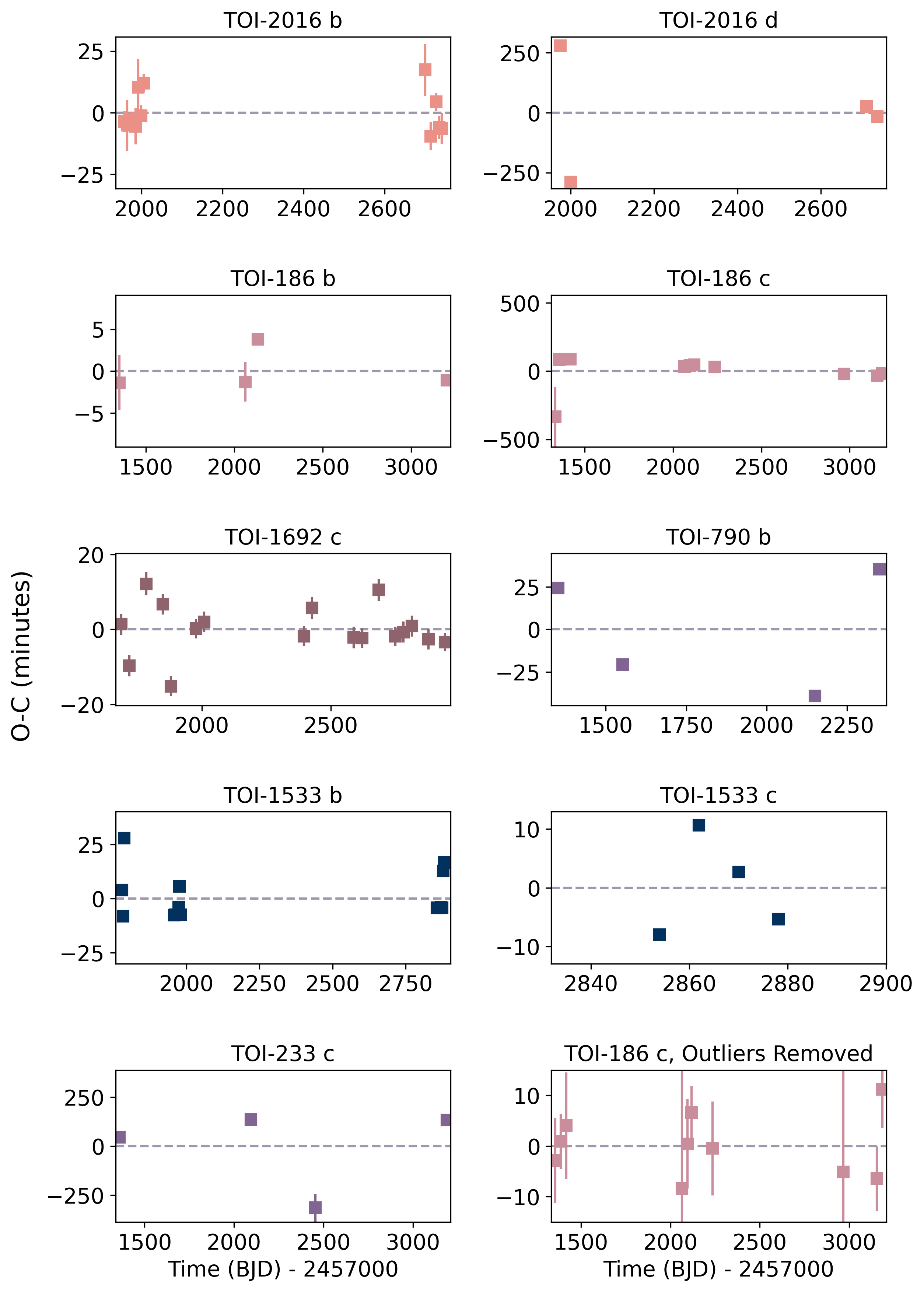}\hfill

\caption{TOIs with significant TTVs in host systems where TTVs had not been previously measured. Continued from Figure \ref{fig:new_ttvs1}. An additional panel is shown for TOI-186\,c with per-sector outlier epochs removed to show underlying structure. Without the outlier, the significance of TOI-186\,c's TTVs diminishes. Nevertheless, these systems represent promising targets for further dynamical follow-up to confirm the presence and nature of their TTVs.}
\label{fig:new_ttvs2}

\end{figure*}

Moreover, for some of these planets we assign transit centers that span a whole sector, so that we can phase-fold transits to increase SNR for smaller planets. We specify which planets make use of per-sector TTV analysis rather than per-transit TTV analysis below. The statistically-significant TTV signals are summarized in Table \ref{tab:newttvs}, as well as plotted in Figures \ref{fig:new_ttvs1} and \ref{fig:new_ttvs2}.

\textbf{\textit{HD 63433/TOI-1726:}} HD 63433 ($V = 6.9$) hosts two super-Earths \citep{Mann:2020} and an Earth-sized planet \citep{Capistrant:2024} in the 400 Myr Ursa Major moving group. Super-Earths HD 63433 b ($P = 7.1079475$ days) and c ($P = 20.543828$ days) were reported as TOI-1726.01 and TOI-1726.02, while the Earth-sized planet HD 63433\,d was not a TOI. Our analysis utilized 5 sectors of data from December 2019 to January 2022, with 18 transits of HD 63433\,b and 8 transits of HD 63433\,c. We find $\sim$12-minute amplitude TTVs for HD 63433\,b and $\sim$5-minute amplitude TTVs for HD 63433\,c (Figure \ref{fig:new_ttvs1}). The observed transit times deviate from the linear ephemeris model by 60$\sigma$ and 13$\sigma$ for HD 63433\,b and HD 63433\,c, respectively. \citet{LopezMurillo:2026} also found $\sim$15-minute TTVs for HD 63433\,b in their independent analysis, with a similar lack of coherent sinusoidal TTV behavior. However, this work is the first to report preliminary TTVs for HD 63433\,c. We also caution that due to the obvious stellar activity in the raw light curves, we cannot rule out that star spots may be inducing TTVs in this particular case. This is because star spots can induce TTVs of a few minutes \citep{Ioannidis:2016}, which are similar to the amplitudes observed here. Therefore, further studies are needed to rule out stellar variability as the cause of the observed TTVs. Stability constraints do allow for the existence of a planet in between the two discovered \citep{Horner:2025}, which exemplifies the need to confirm and characterize potential TTVs in this system.

\textbf{\textit{TOI-4600:}} This system was first identified by \citet{Mireles:2023}, finding two long-period giant planets, TOI-4600\,b ($R = 6.80$\rearth, $P = 82.6869$ days) and TOI-4600\,c ($R = 9.42$\rearth, $P = 482.8191$ days). However, no TTVs had been measured for this system in previous literature. Using seven transits of TOI-4600\,b, spanning 27 sectors from July 2019 to January 2023, we find 10-minute amplitude TTVs (Figure \ref{fig:new_ttvs1}). We also measure a tentative super-period of $\sim$400 days. The measured TTVs of TOI-4600\,b deviate from the linear model by 38$\sigma$. This may suggest the presence of an unseen planet, as TOI-4600\,b and c are not near a first-order resonance. Further analysis will be required to constrain the dynamics of this system, as there are insufficient epochs to examine potential TTVs for TOI-4600\,c.

\textbf{\textit{TOI-772:}} TOI-772 ($R = 0.82 R_\odot$, $M = 0.86 M_\odot$, $V = 11.566$) hosts three planet candidates: sub-Saturns TOI-772.01 ($R = 6.9 R_{\oplus}$, $P \approx 11.02$ days) and TOI-772.02 ($R = 5.3 R_{\oplus}$, $P \approx 744.20$ days), and super-Earth TOI-772.03 ($R = 2.6 R_{\oplus}$, $P \approx 2.79$ days). Due to insufficient epochs for TOI-772.02 and insufficient SNR for TOI-772.03, we are only able to search for TTVs of TOI-772.01. Using six transits of TOI-772.01 across four years, we find 5 minute amplitude TTVs that deviate from a linear ephemeris model by 22$\sigma$. The signal has a tentatively-sinusoidal shape, with a predicted super-period of 3000 days (Figure \ref{fig:new_ttvs1}).

\textbf{\textit{HD 110067:}} This system hosts six sub-Neptunes that were first characterized by \citet{Luque:2023}. A TTV search was carried out, but none were found for the inner three planets (HD 110067 b-d). In our catalog, using six transits of HD 110067\,b ($P = 9.11368$ days) and four transits of HD 110067\,c ($P = 13.673694$ days), we find 6-minute TTVs for HD 110067\,c (Figure \ref{fig:new_ttvs1}). These signals deviate from the linear model by 28$\sigma$ but do not show any coherent sinusoidal behavior. Notably, HD 110067\,b and c lie just wide of the 2:3 resonance, with a normalized distance to the 2:3 MMR of {$\Delta \approx 0.00023$} \citep{Lithwick2012}. We do not measure any TTVs for HD 110067\,b, but the proximity to resonance suggests that TTVs may be detectable with higher-precision instruments. The outer four planets did not have sufficient transits to search for any TTVs.

\textbf{\textit{TOI-4495:}} \textit{TESS} identified two super-Earth sized planet candidates around TOI-4495. TOI-4495.01 was statistically validated in \citet{Hord:2024}, and a TTV search was independently carried out by \citet{Naponiello:2025}. Using 23 transits of TOI-4495.01 (hereafter TOI-4495\,b; $P = 5.183004$ days, $R = 3.63R_\oplus$) and five per-sector transits of TOI-4495.02 (hereafter TOI-4495\,c; $P = 2.569373$ days, $R = 2.18R_\oplus$) across three years, we find 20-minute amplitude TTVs for both planets. This is to be expected, as the planets lie just wide of the 2:1 MMR ($\Delta \approx 0.0086$). The TTVs for TOI-4495\,b and TOI-4495\,c deviate from a flat line model by 6$\sigma$ and 2$\sigma$, respectively (Figure \ref{fig:new_ttvs1}). It should be noted that the significance of the TTVs for TOI-4495\,c increases to 4$\sigma$ when the first sector (with unusually high error) is excluded. Additional epochs will be needed to confirm the significance of the tentative TTVs of TOI-4495\,c. Moreover, we can speculate on the super-period of TOI-4495\,b's TTVs using the observed periodogram peaks above the 0.01\% FAP threshold (Figure \ref{fig:periodograms}). The periodogram shows peaks at around 200 days, which is the same order of magnitude as the predicted super period of $\sim300$ days using Equation 5 from \citet{Lithwick2012}. The peak at $\sim5$ days corresponds to the planet's period. More complete sampling of the TTV signal will increase the strength of the peaks and better constrain the true super-period.

\textbf{\textit{TOI-712:}} This young system of three mini-Neptunes was first characterized by \citet{Vach:2022}. No TTV search was previously performed. With additional baseline, we report a 7-$\sigma$ detection of TTVs for TOI-712\,d ($P = 84.8396$ days, $R = 2.47R_\oplus$), with an amplitude of 40 minutes (Figure \ref{fig:new_ttvs1}). We also find marginal TTVs for TOI-712\,b ($P = 9.531361$ days) and c ($P = 51.69906$ days), but these cannot be confirmed with \textit{TESS} alone at this time. This is because the \textit{TESS} light curves show obvious stellar activity, and in the particular cases of TOI-712\,b and c, we are unable to determine if the TTVs are caused by star spots. In contrast, the 40-minute amplitude TTVs of TOI-712\,d exceed the range of what star spots can induce, which is at most a few minutes \citep{Ioannidis:2016}. Further analysis is crucial to confirm the cause of the measured TTVs, whether that is through a potential 5:3 resonance between TOI-712\,d and c, or due to additional planets.

\begin{figure}[htp]
\centering
\includegraphics[width=.32\textwidth]{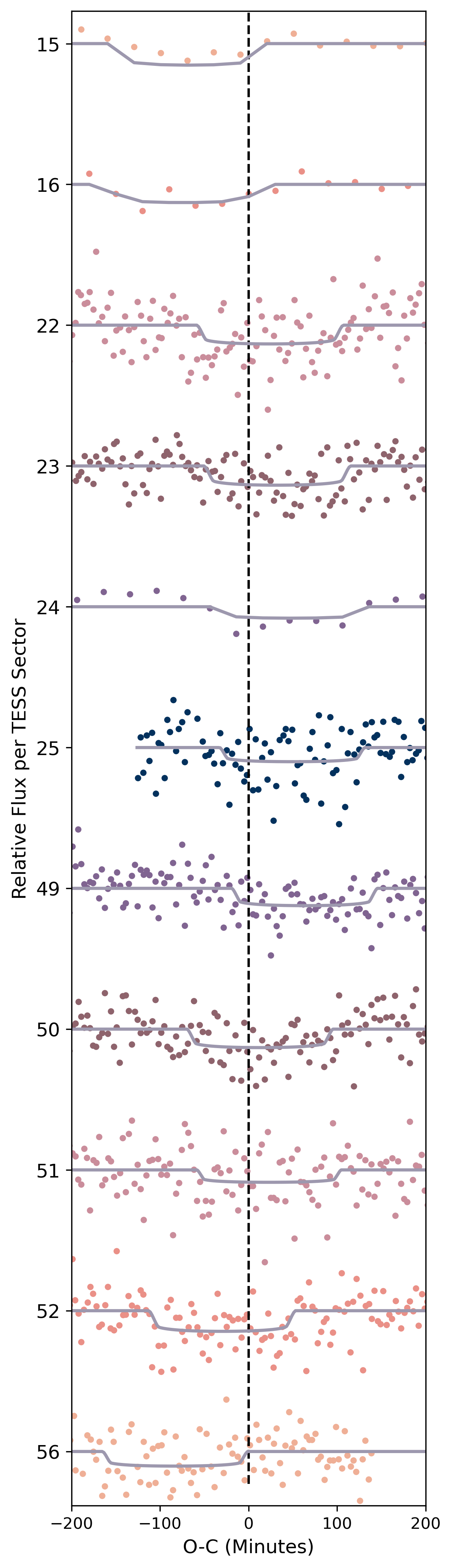}\hfill

\caption{Per-sector TTVs of TOI-1812\,c. The phase-folded, least-squares detrended flux for each sector is plotted with a {\tt batman} model overlaid. A trivial vertical offset is applied for ease of viewing. Differing colors are used to distinguish between neighboring sectors. The x-axis offset shows the observed minus calculated transit times, assuming a linear ephemeris.}
\label{fig:toi1812}

\end{figure}

\textbf{\textit{TOI-1812:}} The TOI catalog reports three TOIs hosted by TOI-1812. Of these, we were only able to investigate the TTVs of the super-Earth TOI-1812.02 (hereafter TOI-1812\,c; $P \approx 11.61$ days, $R = 2.74R_\oplus$); TOI-1812.01 only had one transit observed with \textit{TESS}, and TOI-1812.03 does not have sufficient SNR to detect TTVs. While a full TTV analysis will be published in Osborn et al. (in prep.), we independently find $\sim$1 hour TTVs for TOI-1812\,c using 11 per-sector transits (Figure \ref{fig:toi1812}). These deviate from a linear ephemeris model by 13$\sigma$, with an estimated super-period on the order of 1000 days. This estimation comes from the observed tentative peaks at around 1000 days in the periodogram of TOI-1812\,c (Figure \ref{fig:periodograms}), which lie above the 0.01\% FAP threshold. This is also a consistent order of magnitude with the visual trend observed in the TTV signal (Figure \ref{fig:new_ttvs1}). We are unable to estimate the super-period using Equation 5 from \citet{Lithwick2012} here, as this equation only applies to first-order resonances, and the two planets with measured orbital periods have a $\sim$4:1 orbital period ratio. A more precise constraint of the super-period will require further observed transit epochs.

\textbf{\textit{TOI-2016:}} This system hosts three super-Earth sized TOIs ($R = 2.8-3.3R_\oplus$). Using 13 transits of TOI-2016.01 ($P \approx 6.82$ days) and 4 transits of TOI-2016.03 ($P \approx 25.34$ days), we find 10-minute TTVs for TOI-2016.01 and 5-hour TTVs for TOI-2016.03 (Figure \ref{fig:new_ttvs2}). These signals deviate from the linear model by 3$\sigma$ and 20$\sigma$, respectively. TOI-2016.02 ($P \approx 2.46$ days) does not exhibit discernible TTVs with \textit{TESS} alone. It is unlikely that the observed TTVs are due to interactions only between TOI-2016.01 and TOI-2016.03, as they are not near an MMR, and there is no discernable resonance chain between the three planets.

\textbf{\textit{GJ 143/TOI-186:}} No TTVs were evaluated previously due to insufficient baseline \citep{Dragomir:2019, Trifonov:2019}. However, with additional \textit{TESS} sectors, we find 4 minute TTVs for TOI-186\,b (GJ 143\,b; $P \approx 35.61$ days, $R = 2.61R_\oplus$), and $\sim$100 minute TTVs for TOI-186\,c (HD 21749\,c; $P \approx 7.79$ days, $R = 0.89R_\oplus$). This is measured using four transits of TOI-186\,b and nine per-sector epochs of TOI-186\,c. We report an under-estimated TTV amplitude for TOI-186\,c, as the first sector has lower SNR due to systematics and therefore may be an outlier point. The measured TTVs deviate from a linear model by 2$\sigma$ and 8$\sigma$ for TOI-186\,b and c, respectively. Without the outlier point, the significance of the TTV detection drops to 2$\sigma$ due to its low SNR in \textit{TESS} (Figure \ref{fig:new_ttvs2}). Nevertheless, we report this tentative detection to encourage future follow-up efforts to confirm the nature of these TTVs.

\textbf{\textit{TOI-1692:}} This system hosts two TOIs near the 2:1 resonance. Using 17 transits of TOI-1692\,c ($P = 32.207970$ days, $R = 7.32 R_\oplus$), we find 15  minute TTVs with 7$\sigma$ significance. We are unable to constrain the TTVs of TOI-1692\,b ($P = 17.728661$ days) due to large errorbars, therefore future follow-up will be crucial to confirm the presence of any signals. This is especially useful, as TOI-1692's planets lie narrow of the 2:1 MMR ($\Delta \approx -0.092$).

\textbf{\textit{TOI-790:}} This system hosts three TOIs. Currently, only the eccentric sub-Saturn TOI-790.01 ($P = 199.57791$ days, $R = 6.99 R_\oplus$; hereafter TOI-790\,b) has sufficient epochs and SNR for TTV analysis. TOI-790.03 was only observed in a single transit, and fitting per-sector transit centers for TOI-790.02 ($P = 41.01752$ days) would not increase the SNR so the TTV solution would be unreliable. Using 4 transits of TOI-790\,b, we find 36 minute TTVs with 14$\sigma$ significance and a predicted super-period of 2500 days. TOI-790\,b and TOI-790.02 lie somewhat near the 5:1 resonance, but investigating this higher-order MMR is beyond the scope of this catalog.

\textbf{\textit{TOI-1533:}} This system hosts a sub-Neptune, TOI-1533.01 ($P \approx 3.65$ days, $R = 2.93 R_\oplus$; TOI-1533\,b hereafter), and a Saturn-sized planet candidate, TOI-1533.02 ($P \approx 8.06$ days, $R = 9.29R_\oplus$; TOI-1533\,c hereafter). Using 17 transits of TOI-1533\,b and 4 transits of TOI-1533\,c, we find 30-minute and 10-minute TTVs, respectively. These signals show tentatively-sinusoidal behavior, and they deviate from a linear model by 101$\sigma$ and 36$\sigma$ for planets b and c, respectively. It should be noted that TOI-1533\,c has a V-shaped transit, and consequently we exclude the 30-minute cadence data since the transit duration is 0.8 hours. 

\textbf{\textit{TOI-233:}} This system hosts two super-Earth TOIs. We find 2-hour amplitude TTVs for TOI-233.02 ($P = 7.201138$ days, $R = 1.71R_\oplus$; TOI-233\,c hereafter) using four per-sector transit epochs. This signal deviates from a linear ephemeris by 13$\sigma$. We note here that the Sector 42 $t_0$ is excluded due to insufficient SNR. This is because at least two transits are within the data gap, and one is on the edge of the gap, meaning that any $t_0$ solution for Sector 42 would be unreliable. We do not find any TTVs for TOI-233.01 ($P = 11.670029$ days, $R = 2.02R_\oplus$). There is a potential 5:3 resonance, but future studies will need to confirm this and characterize the proximity to this second-order MMR.

We have 11 systems that have marginal detections with a $\sigma \leq 4$ deviation from a flat line, but we do not list them here as the TTV signal is neither significant nor coherent enough. We cannot rule out all astrophysical false positive scenarios for these systems using \textit{TESS} data alone, and it is beyond the scope of this catalog to do in-depth confirmation of the nature of these TTV signals. It will nonetheless be useful to study them with additional sectors of \textit{TESS} data to further characterize their dynamics.

\begin{deluxetable*}{ccccc}
\tablecaption{
  Newly-Detected TTV Systems
  \label{tab:newttvs}
}
\tablehead{ 
    \colhead{Host Star} & \colhead{Planet} & \colhead{Orbital Period [d]} & \colhead{Approximate TTV Amplitude [min]} & \colhead{Detection Significance}
}
\startdata
HD 63433 & b & 7.1079342 $\pm$ .0000015 & 12 & 60$\sigma$ \\
HD 63433 & c & 20.5438161 $\pm$ .0000054 & 5 & 13$\sigma$ \\
TOI-4600 & b & 82.69013 $\pm$ 0.00020 & 10 & 38$\sigma$ \\
TOI-772 & b & 11.0163419 $\pm$ .0000022 & 5 & 22$\sigma$ \\
HD 110067 & c & 13.6736945 $\pm$ .0000023 & 6 & 28$\sigma$ \\
TOI-4495 & b & 5.183000 $\pm$ .000039 & 20 & 6$\sigma$ \\
TOI-4495 & c & 2.569366 $\pm$ .000024 & 20 & 2$\sigma$\\
TOI-712 & d & 84.83872 $\pm$ 0.00038 & 40 & 7$\sigma$\\
TOI-1812 & c & 11.609806 $\pm$ .000045 & 60 & 13$\sigma$ \\
TOI-2016 & b & 6.816123 $\pm$ .000021 & 10 & 3$\sigma$ \\
TOI-2016 & d & 25.33691 $\pm$ 0.00039 & 300 & 20$\sigma$ \\
TOI-186 & b & 35.613436 $\pm$ .000028 & 4 & 2$\sigma$\\
TOI-186 & c & 7.789773 $\pm$ .000021 & 100 & 8$\sigma$\\
TOI-1692 & c & 32.207967 $\pm$ .000018 & 15 & 7$\sigma$ \\
TOI-790 & b & 199.57782 $\pm$ 0.00090 & 36 & 14$\sigma$\\
TOI-1533 & b & 3.6458045 $\pm$ .0000014 & 30 & 101$\sigma$\\
TOI-1533 & c & 8.0637952 $\pm$ .0000022 & 10 & 36$\sigma$\\
TOI-233 & c & 7.201137 $\pm$ .000014 & 120 & 13$\sigma$\\
\enddata
\end{deluxetable*}

\subsubsection{Systems with TTVs that are not detectable in TESS alone}

In some cases, \textit{TESS} data alone has insufficient SNR to reliably detect TTVs. Therefore, there are 9 systems with known TTVs that cannot be detected in this catalog. This is due to a number of factors, from low per-transit SNR to observational gaps. The following systems have confirmed TTVs whose signals were unrecoverable using \textit{TESS} alone:

\begin{itemize}
    \item \textbf{TOI-5398:} Our results are consistent with the findings in \citet{Mantovan:2023}, which conclude that TTVs may be present in the system but are unconstrained using only \textit{TESS} data. Even with additional \textit{TESS} data, this finding remains unchanged. Further follow-up analysis using multi-instrument photometry is needed to confirm or deny the presence of TTVs.
    
    \item \textbf{TOI-2076:} TTVs were first characterized in \citet{Osborn:2022}, reporting 10-minute TTVs for planets b and d, and 20 minute TTVs for planet c. However, this analysis primarily used a combination of CHEOPS and ground-based data, with \textit{TESS} data as a supplement. These additional facilities were instrumental in detecting TTVs, as they sampled more of the higher-amplitude portion of the TTV supercycle. In this analysis, we are not able to confirm the TTV signal using \textit{TESS} alone due to insufficient SNR.
    
    \item \textbf{Kepler-89/TOI-4581:} Kepler-89 hosts four planets with measured TTVs using \textit{Kepler} data. The published TTV amplitudes are approximately 130 minutes (Kepler-89\,b), 10 minutes (Kepler-89\,c), 2 minutes (Kepler-89\,d), and 7 minutes (Kepler-89\,e) \citep{Weiss:2013}. Only Kepler-89\,d and e were observed in \textit{TESS}, with only one transit of Kepler-89\,e having been captured. Therefore, the dearth of transit observations and low semi-amplitudes mean that the current \textit{TESS} data is insufficient to reproduce previously-published TTV results for Kepler-89.
    
    \item \textbf{K2-266/TOI-5520:} K2-266 is a four-planet system with published TTVs, as K2-266\,d and e are near the 4:3 MMR. Due to this proximity to resonance, K2-266\,d and e have measured TTVs of approximately 15 minutes in amplitude \citep{Rodriguez:2018}. However, while K2-266\,d and e are observed in \textit{TESS}, the per-transit errorbars of $\sim$1 hour preclude detection of their TTVs.
    
    \item \textbf{TOI-836:} This system hosts two transiting planets, TOI-836\,b ($P \approx 3.82$ days) and TOI-836\,c ($P \approx 8.6$ days). TOI-836\,c has TTVs of $\sim$20 minutes \citep{Hawthorn:2023}, although this was determined using additional transit observations from CHEOPS and LCOGT. Consistent with our own findings, the \textit{TESS} points alone from \citet{Hawthorn:2023} did not show sufficient deviation from a linear ephemeris, as the CHEOPS and LCOGT transits sampled the peaks of the TTV signal. Therefore, we were unable to recover the TTVs of TOI-836\,c through sole use of \textit{TESS} data.
    
    \item \textbf{TOI-5126:} Tentative (2$\sigma$) TTVs were suggested in \citet{Fairnington:2024}, however additional transits were needed to rule out any false positive scenarios and confirm the signals. We find similar results, whereby the \textit{TESS} data lacks sufficient SNR to confirm any TTVs at this time.
    
    \item \textbf{TOI-1803:} This system hosts two sub-Neptune planets near the 2:1 MMR. Using combined photometric data from \textit{TESS}, CHEOPS, and LCOGT, \citet{Zingales:2025} find TTVs of 10 minutes for TOI-1803/,b and 40 minutes for TOI-1803\,c. However, the \textit{TESS} transits alone were consistent with a linear ephemeris, as the LCOGT and CHEOPS datapoints were crucial for the solid detection of the system's TTVs. In our \textit{TESS}-only analysis, we find that the TTVs are unconstrained, as was the case in previous work.
    
    \item \textbf{TOI-270:} TOI-270 hosts a super-Earth (TOI-270\,b) and two sub-Neptunes (TOI-270\,c and d). Upon the initial discovery of the system by \citet{Guenther:2019}, the authors also agree with our result that TTVs are not shown in \textit{TESS}. We also concur with their suggestion that TTVs could nevertheless exist due to proximity of TOI-270\,c and d to the 2:1 resonance. Further high-precision transit follow-up will be necessary to confirm the TTVs of TOI-270's planets.
    
    \item \textbf{TOI-1266:} Previous TTV searches of TOI-1266 were conducted by \citet{Demory:2020} and \citet{Greklek:2025}. \citet{Demory:2020} reported a preliminary TTV detection for TOI-1266\,b and c, and \citet{Greklek:2025} performed an updated TTV analysis with extended baseline. In the latter work, a third non-transiting planet was confirmed through TTV modeling. In our search, the TTVs of TOI-1266\,b from \textit{TESS} match those in \citet{Greklek:2025}, and we are similarly unable to recover the TTVs of TOI-1266\,c using \textit{TESS} alone.
    
\end{itemize}

\subsection{Comparison with Pre-\textit{TESS} TTV Simulations}

We compare our TTV yield to that predicted by the \citet{Hadden:2019} coverage configuration with Camera 3 centered on the ecliptic pole and a 3-year extended mission with observations beginning in the north (E$_{3,NSN}$). This is the closest model to the true \textit{TESS} observing configuration, which had Camera 3 centered on the ecliptic pole but started in the south with alternating hemisphere pointing sequences. We caveat that the observations are not exactly the same as the yield simulation. The real pointing sequence led to less continuous coverage but more stars observed. Our SNR threshold for target selection is also higher than assumed in the simulation, which may lead to fewer TTV systems analyzed and recovered. 

Table \ref{tab:yields} compares the multiplicity statistics and TTV yields in this work to those predicted by \citet{Hadden:2019}, where we consider both of their predicted fiducial ($\sigma_i = 2^\circ$) and ``low-$i$" ($\sigma_i = 1.5^\circ$) solutions. These solutions model the inclination distribution of \textit{TESS} systems as a Gaussian with $\sigma_i = 2^{\circ}$ or $\sigma_i = 1.5^{\circ}$.

In their synthetic fiducial population, \citet{Hadden:2019} predicted that 25 multi-transiting planet systems would be found to host significant TTVs during the \textit{TESS} Extended Mission (defined as Category 1 and 2 in Table \ref{tab:yields}). Indeed, our catalog finds a consistent observational result, with 20 systems exhibiting significant TTVs. Moreover, by the end of the third \textit{TESS} Extended Mission, the additional baseline will likely confirm the now-marginal detections that we do not present here. 

Our catalog demonstrates the potential of TTVs in characterizing the architectures of \textit{TESS} systems as a whole. Our observed TTV yield begins to show preferential support for the $\sigma_i = 2^{\circ}$ solution compared to the low-$i$ solution. 

Future studies should perform updated simulations with real \textit{TESS} data to help constrain the planetary mutual inclination distribution. Additional TTV systems are needed to generate a statistically significant result, but this catalog is nonetheless a promising proof of concept that uniform TTV catalogs are useful tools in understanding the mutual inclination distribution. As \textit{TESS} continues to identify multi-planet systems, predicted TTV yields must also be updated to help rule out theoretical mutual inclination solutions.

We also note a significantly higher percentage of 3- and 4-planet systems in the sample we have, indicating a potential mismatch in the true versus simulated period spacing distribution. The higher percentage of 3- and 4-planet systems observed indicates that the physical spacing of \textit{TESS} planets may be more compact than originally expected from the \citet{Hadden:2019} simulation. This is because more compact systems are more likely to have higher transiting planet multiplicity. This could also be due to the differences in the \textit{TESS} observing strategy employed in the simulation versus the actual mission.

This catalog expands upon the statistical sample produced by the previous \textit{TESS} TTV catalog by \citet{Naponiello:2025} by including systems with multiple transiting planet candidates. This probes more compact systems in contrast to the previous catalog's search for non-transiting companions. The previous catalog also mandates that each system hosts at least one confirmed transiting planet. In this work, we add 13 new TTV systems, which is a $\sim$60\% increase in the number of \textit{TESS} systems with significant TTVs (as of Sector 69).

\begin{deluxetable}{cccccc}
\tablecaption{
  Predicted vs Observed \textit{TESS} Planet Multiplicity \& TTV Yield
  \label{tab:yields}
}
\tablehead{ 
    \colhead{$N$ Transiting} & \colhead{2} & \colhead{3} & \colhead{4} & \multicolumn{2}{c}{{TTV Yield}}
}

\startdata
 & & & & Cat 1 + 2\tablenotemark{a} & Cat 3\\ \hline
\textbf{This work} & 134 & 31 & 5 & 20 & 11 \\
$\sigma_i = 2^\circ$ & 143 & 3 & 1 & 25 & 7 \\
$\sigma_i = 1.5^\circ$ & 166 & 7 & 0 & 40 & 11\\
\enddata
\tablenotetext{a}{The definitions for Category 1, 2, and 3 systems from \citet{Hadden:2019} do not share the exact same significance thresholds as in this work, but are still suitable for rough comparison.}
\end{deluxetable}

\subsection{Sample Breakdown and Comparison to Kepler}

This catalog presents the first snapshot of the population of TTV systems in \textit{TESS} multi-TOI systems. We find that TTV systems in \textit{TESS} have slightly different architectures than those found by \textit{Kepler}, with \textit{TESS} TTV systems exhibiting a pile-up at the 2:1 resonance (Figure \ref{fig:pratiohist}, left). In contrast, the \textit{Kepler} TTV population shows peaks at the 3:2 and 3:1 resonances. This may point to different system architectures in \textit{TESS} versus \textit{Kepler} systems, though this could also be biased due to the shorter duration of \textit{TESS} sectors. Future population synthesis studies will need to examine the completeness of the \textit{TESS} sample to confirm whether this is an astrophysical feature or due to observational bias.

On the other hand, the distribution of neighboring planet radius ratios in \textit{TESS} and \textit{Kepler} TTV systems are roughly consistent (Figure \ref{fig:pratiohist}, right). The populations of TTV systems in \textit{TESS} versus \textit{Kepler} with regards to period-radius space are also consistent (Figure \ref{fig:ttvscatter}). This may be due to the fact that both missions are sensitive to similar populations of compact systems, which are geometrically more likely to result in multiple transiting planets. It is not yet clear whether these findings represent an inherent property of TTV systems, or an observational bias of the transit method.

As TTV strength scales with mass ratio between neighboring planets, we also investigate the fractions of TTV systems with and without giant planets. Of the 19 systems that host giant planets, 6 exhibit TTVs in \textit{TESS} alone ($\sim$32\%). In a similar fashion, 16 out of 57 systems without giant planets host TTVs ($\sim$28\%). This is consistent with the TTV yield from \citet{Holczer:2016}, which found that 6 out of 24 systems (25\%) hosting giant planets had significant long-term TTVs, while 254 out of 504 ($\sim$50\%) of systems with only smaller planets showed significant TTVs. This suggests that proximity to resonance is the driving force in generating detectable TTVs, rather than the mere presence of a giant planet.

\subsection{Limitations and Future Work}

Our results could be skewed to reflect the observational biases that shape the \textit{TESS} sample, such as sensitivity to close-in, large planets. Therefore, future work will require a population-synthesis approach to mitigate these biases and determine the true characteristics of the population of TTV systems.

One other limitation of note is that the current \textit{TESS} baseline does not sufficiently sample the super-period of most systems, as the super-periods of the significant TTV systems here can reach thousands of days. This poor sampling precludes strong peaks in the Lomb-Scargle periodograms (Figure \ref{fig:periodograms}), and highlights the need for continuous follow up of \textit{TESS} systems to measure the precise characteristics of the TTV cycle. 

Future follow-up will also be instrumental in characterizing the new TTV systems presented in this work. In-depth dynamical analyses with longer baseline and ground-based transit observations will help confirm both the planetary nature of planet candidates in the catalog, as well as the cause of the observed TTVs. Sampling the TTV signal with ground-based observations is particularly advantageous, and this catalog presents many promising targets for ground-based campaigns to fully characterize the TTVs identified here.

\section*{Acknowledgements}
E.N. acknowledges the PhD scholarship provided by the Australian Research Council discovery grant DP220100365. C.X.H acknowledges that her research is sponsored by the Australian Research Council DECRA Fellowship DE200101840 and Future Fellowship FT240100016. G.Z. acknowledges that his research is sponsored by the Australian Research Council Future Fellowship FT230100517.


This research has made use of the Exoplanet Follow-up Observation Program \citep[EXOFOP;][]{https://doi.org/10.26134/exofop5} website, which is operated by the California Institute of Technology, under contract with the National Aeronautics and Space Administration under the Exoplanet Exploration Program.


Funding for the TESS mission is provided by NASA's Science Mission Directorate.

\textit{Software:} {\tt batman} \citep{batman}, {\tt emcee} \citep{Foreman-Mackey2013}, {\tt matplotlib} \citep{Hunter2007}, {\tt pandas} \citep{McKinney2010}, {\tt scipy} \citep{Virtanen:2020}. 

\facilities{\textit{TESS}; NASA Exoplanet Archive.}


\appendix
\section{Lomb-Scargle Periodograms of TTV Signals}
\label{appendix}

\begin{figure*}[htp]
\centering
\includegraphics[width=.9\textwidth]{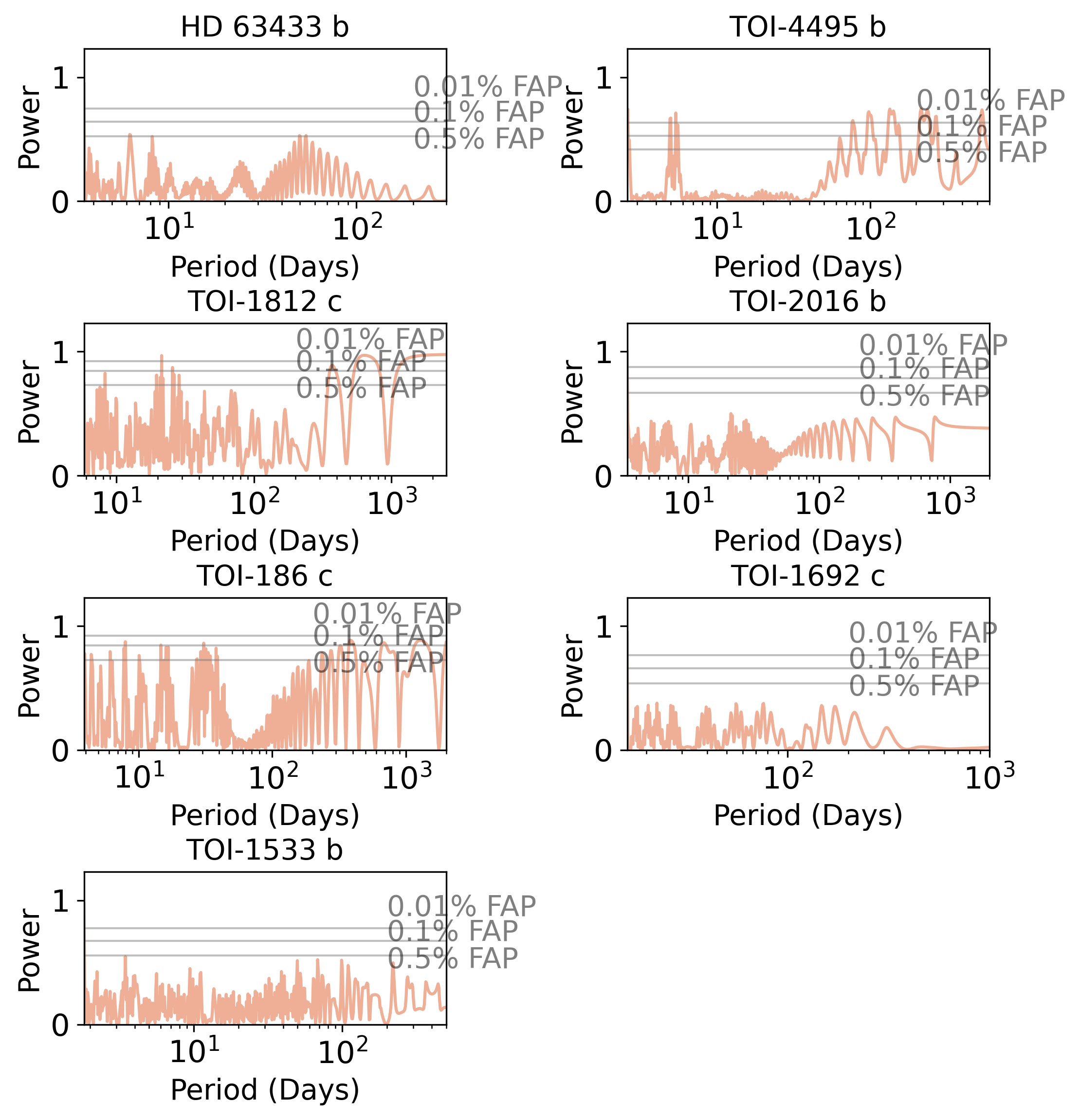}\hfill

\caption{Lomb-Scargle periodograms of TTV signals in systems with more than ten transits observed. The 0.01\%, 0.1\%, and 0.5\% False Alarm Probabilities are labeled with corresponding horizontal lines. There are no strong peaks in the periodograms due to the expected TTV super-periods far exceeding the current baseline of \textit{TESS}. Measurement of the super-periods will thus be left to future work as the \textit{TESS} baseline increases.}
\label{fig:periodograms}

\end{figure*}

\section{Best-Fit Transit Times for Significant TOIs}

\startlongtable
\begin{deluxetable}{cccccc}
\tablecaption{Best-Fit Individual Transit Times for Significant TTV Planets \label{tab:ttvfull}}

\tablehead{
\colhead{TIC} & \colhead{Planet} & \colhead{Epoch} & \colhead{$T_0$ [BJD]} & \colhead{$\sigma_{T_0}$ [d]} & \colhead{Per-Sector Flag}
}

\startdata
55652896 & TOI-216.01 & 0 & 2458331.28578 & 0.00071 & N \\
55652896 & TOI-216.01 & 1 & 2458365.82401 & 0.00072 & N \\
55652896 & TOI-216.01 & 2 & 2458400.36890 & 0.00073 & N \\
55652896 & TOI-216.01 & 3 & 2458434.92225 & 0.00071 & N \\
55652896 & TOI-216.01 & 4 & 2458469.47740 & 0.00072 & N \\
55652896 & TOI-216.01 & 6 & 2458538.59241 & 0.00069 & N \\
55652896 & TOI-216.01 & 8 & 2458607.70771 & 0.00074 & N \\
55652896 & TOI-216.01 & 10 & 2458676.80889 & 0.00087 & N \\
55652896 & TOI-216.01 & 21 & 2459056.35289 & 0.00081 & N \\
55652896 & TOI-216.01 & 22 & 2459090.81125 & 0.00066 & N \\
55652896 & TOI-216.01 & 23 & 2459125.27059 & 0.00082 & N \\
55652896 & TOI-216.01 & 24 & 2459159.72046 & 0.00074 & N \\
55652896 & TOI-216.01 & 25 & 2459194.16704 & 0.00066 & N \\
55652896 & TOI-216.01 & 27 & 2459263.06026 & 0.00076 & N \\
55652896 & TOI-216.01 & 28 & 2459297.51121 & 0.00089 & N \\
55652896 & TOI-216.01 & 29 & 2459331.95899 & 0.00072 & N \\
55652896 & TOI-216.01 & 30 & 2459366.40924 & 0.00074 & N \\
55652896 & TOI-216.01 & 49 & 2460022.20733 & 0.00081 & N \\
55652896 & TOI-216.01 & 50 & 2460056.75970 & 0.00081 & N \\
55652896 & TOI-216.01 & 51 & 2460091.30996 & 0.00078 & N \\
55652896 & TOI-216.01 & 53 & 2460160.39545 & 0.00011 & N \\
55652896 & TOI-216.02 & 1 & 2458342.43550 & 0.00099 & N \\
55652896 & TOI-216.02 & 2 & 2458359.5422 & 0.0025 & N \\
55652896 & TOI-216.02 & 3 & 2458376.6290 & 0.0024 & N \\
55652896 & TOI-216.02 & 4 & 2458393.7225 & 0.0023 & N \\
55652896 & TOI-216.02 & 6 & 2458427.8783 & 0.0025 & N \\
55652896 & TOI-216.02 & 7 & 2458444.9607 & 0.0024 & N \\
55652896 & TOI-216.02 & 8 & 2458462.0317 & 0.0019 & N \\
55652896 & TOI-216.02 & 9 & 2458479.0918 & 0.0024 & N \\
55652896 & TOI-216.02 & 10 & 2458496.1562 & 0.0025 & N \\
55652896 & TOI-216.02 & 11 & 2458513.2298 & 0.0021 & N \\
55652896 & TOI-216.02 & 13 & 2458547.3371 & 0.0024 & N \\
55652896 & TOI-216.02 & 14 & 2458564.4054 & 0.0027 & N \\
55652896 & TOI-216.02 & 18 & 2458632.6738 & 0.0033 & N \\
55652896 & TOI-216.02 & 19 & 2458649.7570 & 0.0026 & N \\
55652896 & TOI-216.02 & 20 & 2458666.8581 & 0.0028 & N \\
55652896 & TOI-216.02 & 42 & 2459045.4623 & 0.0022 & N \\
55652896 & TOI-216.02 & 43 & 2459062.8048 & 0.0015 & N \\
55652896 & TOI-216.02 & 44 & 2459080.1571 & 0.0022 & N \\
55652896 & TOI-216.02 & 45 & 2459097.5091 & 0.0026 & N \\
55652896 & TOI-216.02 & 47 & 2459132.2437 & 0.0018 & N \\
55652896 & TOI-216.02 & 48 & 2459149.6256 & 0.0021 & N \\
55652896 & TOI-216.02 & 49 & 2459167.0063 & 0.0019 & N \\
55652896 & TOI-216.02 & 50 & 2459184.3957 & 0.0022 & N \\
55652896 & TOI-216.02 & 52 & 2459219.1807 & 0.0024 & N \\
55652896 & TOI-216.02 & 53 & 2459236.569 & 0.002 & N \\
55652896 & TOI-216.02 & 54 & 2459253.965 & 0.002 & N \\
55652896 & TOI-216.02 & 56 & 2459288.7530 & 0.0026 & N \\
55652896 & TOI-216.02 & 58 & 2459323.5260 & 0.0023 & N \\
55652896 & TOI-216.02 & 59 & 2459340.9049 & 0.0022 & N \\
55652896 & TOI-216.02 & 60 & 2459358.2833 & 0.0019 & N \\
130181866 & HD 63433.01 & 0 & 2458845.37951 & 0.00017 & N \\
130181866 & HD 63433.01 & 1 & 2458852.48100 & 0.00021 & N \\
130181866 & HD 63433.01 & 2 & 2458859.5889 & 0.0002 & N \\
130181866 & HD 63433.01 & 3 & 2458866.6923 & 0.0002 & N \\
130181866 & HD 63433.01 & 93 & 2459506.41095 & 0.00018 & N \\
130181866 & HD 63433.01 & 95 & 2459520.62677 & 0.00021 & N \\
130181866 & HD 63433.01 & 96 & 2459527.73984 & 0.00018 & N \\
130181866 & HD 63433.01 & 97 & 2459534.84263 & 0.00021 & N \\
130181866 & HD 63433.01 & 98 & 2459541.9599 & 0.0002 & N \\
130181866 & HD 63433.01 & 99 & 2459549.06985 & 0.00019 & N \\
130181866 & HD 63433.01 & 100 & 2459556.1678 & 0.0002 & N \\
130181866 & HD 63433.01 & 101 & 2459563.27435 & 0.00019 & N \\
130181866 & HD 63433.01 & 102 & 2459570.38230 & 0.00023 & N \\
130181866 & HD 63433.01 & 103 & 2459577.49026 & 0.00015 & N \\
130181866 & HD 63433.01 & 104 & 2459584.5982 & 0.0002 & N \\
130181866 & HD 63433.01 & 105 & 2459591.7132 & 0.0002 & N \\
130181866 & HD 63433.01 & 106 & 2459598.81940 & 0.00022 & N \\
130181866 & HD 63433.01 & 107 & 2459605.92200 & 0.00017 & N \\
130181866 & HD 63433.02 & -36 & 2458844.05974 & 0.00021 & N \\
130181866 & HD 63433.02 & -35 & 2458864.6048 & 0.0002 & N \\
130181866 & HD 63433.02 & -4 & 2459501.45808 & 0.00018 & N \\
130181866 & HD 63433.02 & -3 & 2459522.00568 & 0.00017 & N \\
130181866 & HD 63433.02 & -2 & 2459542.54944 & 0.00017 & N \\
130181866 & HD 63433.02 & -1 & 2459563.09330 & 0.00019 & N \\
130181866 & HD 63433.02 & 0 & 2459583.63708 & 0.00021 & N \\
130181866 & HD 63433.02 & 1 & 2459604.1803 & 0.0002 & N \\
149601126 & TOI-2525.01 & 1 & 2458333.5270 & 0.0054 & N \\
149601126 & TOI-2525.01 & 5 & 2458426.7839 & 0.0057 & N \\
149601126 & TOI-2525.01 & 7 & 2458473.5019 & 0.0054 & N \\
149601126 & TOI-2525.01 & 8 & 2458496.8604 & 0.0048 & N \\
149601126 & TOI-2525.01 & 9 & 2458520.2385 & 0.0045 & N \\
149601126 & TOI-2525.01 & 11 & 2458566.9607 & 0.0036 & N \\
149601126 & TOI-2525.01 & 15 & 2458659.9862 & 0.0093 & N \\
149601126 & TOI-2525.01 & 32 & 2459056.0521 & 0.0048 & N \\
149601126 & TOI-2525.01 & 33 & 2459079.2668 & 0.0037 & N \\
149601126 & TOI-2525.01 & 35 & 2459125.6424 & 0.0023 & N \\
149601126 & TOI-2525.01 & 36 & 2459148.8351 & 0.0035 & N \\
149601126 & TOI-2525.01 & 38 & 2459195.310 & 0.003 & N \\
149601126 & TOI-2525.01 & 39 & 2459218.6011 & 0.0045 & N \\
149601126 & TOI-2525.01 & 41 & 2459265.2381 & 0.0029 & N \\
149601126 & TOI-2525.01 & 42 & 2459288.5972 & 0.0033 & N \\
149601126 & TOI-2525.01 & 43 & 2459311.9518 & 0.0027 & N \\
149601126 & TOI-2525.01 & 45 & 2459358.6861 & 0.0045 & N \\
149601126 & TOI-2525.02 & 1 & 2458335.4101 & 0.0087 & N \\
149601126 & TOI-2525.02 & 3 & 2458433.8716 & 0.0047 & N \\
149601126 & TOI-2525.02 & 4 & 2458483.0909 & 0.0034 & N \\
149601126 & TOI-2525.02 & 8 & 2458680.0668 & 0.0052 & N \\
149601126 & TOI-2525.02 & 17 & 2459123.2688 & 0.0023 & N \\
149601126 & TOI-2525.02 & 19 & 2459221.7911 & 0.0027 & N \\
232608943 & TOI-4600.01 & -8 & 2458757.89273 & 0.00017 & N \\
232608943 & TOI-4600.01 & -6 & 2458923.27359 & 0.00017 & N \\
232608943 & TOI-4600.01 & -5 & 2459005.95319 & 0.00018 & N \\
232608943 & TOI-4600.01 & 2 & 2459584.76938 & 0.00018 & N \\
232608943 & TOI-4600.01 & 4 & 2459750.14391 & 0.00022 & N \\
232608943 & TOI-4600.01 & 5 & 2459832.83405 & 0.00018 & N \\
232608943 & TOI-4600.01 & 6 & 2459915.53076 & 0.00024 & N \\
254113311 & TOI-1130.01 & 0 & 2458657.90322 & 0.00017 & N \\
254113311 & TOI-1130.01 & 1 & 2458666.25340 & 0.00021 & N \\
254113311 & TOI-1130.01 & 2 & 2458674.60369 & 0.00025 & N \\
254113311 & TOI-1130.01 & 46 & 2459041.99680 & 0.00016 & N \\
254113311 & TOI-1130.01 & 47 & 2459050.34675 & 0.00017 & N \\
254113311 & TOI-1130.01 & 48 & 2459058.6964 & 0.0002 & N \\
254113311 & TOI-1130.01 & 176 & 2460127.4352 & 0.0002 & N \\
254113311 & TOI-1130.01 & 177 & 2460135.78641 & 0.00021 & N \\
254113311 & TOI-1130.01 & 178 & 2460144.13630 & 0.00022 & N \\
254113311 & TOI-1130.02 & 0 & 2458658.74403 & 0.00023 & N \\
254113311 & TOI-1130.02 & 1 & 2458662.81336 & 0.00021 & N \\
254113311 & TOI-1130.02 & 2 & 2458666.8862 & 0.0002 & N \\
254113311 & TOI-1130.02 & 3 & 2458670.96496 & 0.00019 & N \\
254113311 & TOI-1130.02 & 4 & 2458675.03412 & 0.00023 & N \\
254113311 & TOI-1130.02 & 5 & 2458679.07585 & 0.00021 & N \\
254113311 & TOI-1130.02 & 93 & 2459037.94532 & 0.00016 & N \\
254113311 & TOI-1130.02 & 95 & 2459046.07625 & 0.00021 & N \\
254113311 & TOI-1130.02 & 96 & 2459050.14245 & 0.00023 & N \\
254113311 & TOI-1130.02 & 97 & 2459054.21692 & 0.00021 & N \\
254113311 & TOI-1130.02 & 98 & 2459058.29130 & 0.00024 & N \\
254113311 & TOI-1130.02 & 360 & 2460126.92345 & 0.00021 & N \\
254113311 & TOI-1130.02 & 361 & 2460130.99064 & 0.00021 & N \\
254113311 & TOI-1130.02 & 362 & 2460135.05924 & 0.00019 & N \\
254113311 & TOI-1130.02 & 364 & 2460143.20080 & 0.00026 & N \\
254113311 & TOI-1130.02 & 365 & 2460147.27670 & 0.00017 & N \\
286864983 & TOI-772.01 & 0 & 2458575.9454 & 0.0002 & N \\
286864983 & TOI-772.01 & 1 & 2458586.96416 & 0.00026 & N \\
286864983 & TOI-772.01 & 67 & 2459314.03693 & 0.00024 & N \\
286864983 & TOI-772.01 & 68 & 2459325.05396 & 0.00023 & N \\
286864983 & TOI-772.01 & 134 & 2460052.13597 & 0.00013 & N \\
286864983 & TOI-772.01 & 135 & 2460063.15527 & 0.00016 & N \\
347332255 & HD 110067.01 & -77 & 2458938.40472 & 0.00018 & N \\
347332255 & HD 110067.01 & -76 & 2458947.51844 & 0.00017 & N \\
347332255 & HD 110067.01 & 0 & 2459640.15797 & 0.00022 & N \\
347332255 & HD 110067.01 & 1 & 2459649.27160 & 0.00022 & N \\
347332255 & HD 110067.01 & 2 & 2459658.38536 & 0.00023 & N \\
347332255 & HD 110067.02 & -53 & 2458932.7512 & 0.0002 & N \\
347332255 & HD 110067.02 & -52 & 2458946.41328 & 0.00022 & N \\
347332255 & HD 110067.02 & -1 & 2459643.78335 & 0.00025 & N \\
347332255 & HD 110067.02 & 0 & 2459657.45704 & 0.00017 & N \\
120826158 & TOI-4495.01 & -208 & 2458687.8461 & 0.0036 & N \\
120826158 & TOI-4495.01 & -206 & 2458698.2083 & 0.0037 & N \\
120826158 & TOI-4495.01 & -205 & 2458703.3989 & 0.0066 & N \\
120826158 & TOI-4495.01 & -72 & 2459392.7496 & 0.0038 & N \\
120826158 & TOI-4495.01 & -71 & 2459397.9401 & 0.0042 & N \\
120826158 & TOI-4495.01 & -70 & 2459403.1156 & 0.0047 & N \\
120826158 & TOI-4495.01 & -69 & 2459408.2916 & 0.0029 & N \\
120826158 & TOI-4495.01 & -68 & 2459413.4742 & 0.0024 & N \\
120826158 & TOI-4495.01 & -67 & 2459418.6593 & 0.0064 & N \\
120826158 & TOI-4495.01 & -66 & 2459423.8348 & 0.0029 & N \\
120826158 & TOI-4495.01 & -65 & 2459429.0200 & 0.0023 & N \\
120826158 & TOI-4495.01 & -64 & 2459434.1960 & 0.0029 & N \\
120826158 & TOI-4495.01 & -63 & 2459439.3790 & 0.0053 & N \\
120826158 & TOI-4495.01 & -62 & 2459444.5642 & 0.0033 & N \\
120826158 & TOI-4495.01 & -4 & 2459745.1888 & 0.0051 & N \\
120826158 & TOI-4495.01 & -2 & 2459755.5595 & 0.0033 & N \\
120826158 & TOI-4495.01 & -1 & 2459760.741 & 0.002 & N \\
120826158 & TOI-4495.01 & 0 & 2459765.9267 & 0.0038 & N \\
120826158 & TOI-4495.01 & 1 & 2459771.1101 & 0.0035 & N \\
120826158 & TOI-4495.01 & 2 & 2459776.3012 & 0.0054 & N \\
120826158 & TOI-4495.01 & 3 & 2459781.4751 & 0.0029 & N \\
120826158 & TOI-4495.01 & 4 & 2459786.6630 & 0.0054 & N \\
120826158 & TOI-4495.01 & 5 & 2459791.8486 & 0.0035 & N \\
120826158 & TOI-4495.02 & -422 & 2458685.773 & 0.013 & Y \\
120826158 & TOI-4495.02 & -147 & 2459392.3135 & 0.0021 & Y \\
120826158 & TOI-4495.02 & -136 & 2459420.5964 & 0.0026 & Y \\
120826158 & TOI-4495.02 & -10 & 2459744.3136 & 0.0036 & Y \\
120826158 & TOI-4495.02 & 0 & 2459769.9975 & 0.0044 & Y \\
150151262 & TOI-712.01 & -38 & 2458395.0727 & 0.0045 & N \\
150151262 & TOI-712.01 & -36 & 2458414.1344 & 0.0035 & N \\
150151262 & TOI-712.01 & -30 & 2458471.3253 & 0.0057 & N \\
150151262 & TOI-712.01 & -27 & 2458499.9293 & 0.0046 & N \\
150151262 & TOI-712.01 & -22 & 2458547.5797 & 0.0054 & N \\
150151262 & TOI-712.01 & -19 & 2458576.1746 & 0.0031 & N \\
150151262 & TOI-712.01 & -16 & 2458604.7686 & 0.0052 & N \\
150151262 & TOI-712.01 & -10 & 2458661.9564 & 0.0037 & N \\
150151262 & TOI-712.01 & 30 & 2459043.2060 & 0.0054 & N \\
150151262 & TOI-712.01 & 35 & 2459090.8692 & 0.0043 & N \\
150151262 & TOI-712.01 & 38 & 2459119.4605 & 0.0047 & N \\
150151262 & TOI-712.01 & 41 & 2459148.0583 & 0.0034 & N \\
150151262 & TOI-712.01 & 44 & 2459176.6433 & 0.0044 & N \\
150151262 & TOI-712.01 & 47 & 2459205.2443 & 0.0031 & N \\
150151262 & TOI-712.01 & 50 & 2459233.8360 & 0.0038 & N \\
150151262 & TOI-712.01 & 56 & 2459291.0239 & 0.0076 & N \\
150151262 & TOI-712.01 & 58 & 2459310.0899 & 0.0045 & N \\
150151262 & TOI-712.01 & 64 & 2459367.2783 & 0.0045 & N \\
150151262 & TOI-712.01 & 127 & 2459967.7535 & 0.0026 & N \\
150151262 & TOI-712.01 & 130 & 2459996.3467 & 0.0045 & N \\
150151262 & TOI-712.01 & 132 & 2460015.4060 & 0.0039 & N \\
150151262 & TOI-712.01 & 135 & 2460044.0024 & 0.0024 & N \\
150151262 & TOI-712.01 & 138 & 2460072.5994 & 0.0022 & N \\
150151262 & TOI-712.01 & 141 & 2460101.194 & 0.004 & N \\
150151262 & TOI-712.01 & 144 & 2460129.7832 & 0.0043 & N \\
150151262 & TOI-712.02 & -10 & 2458429.4056 & 0.0072 & N \\
150151262 & TOI-712.02 & -9 & 2458481.1101 & 0.0079 & N \\
150151262 & TOI-712.02 & 2 & 2459049.8048 & 0.0062 & N \\
150151262 & TOI-712.02 & 4 & 2459153.1924 & 0.0087 & N \\
150151262 & TOI-712.02 & 5 & 2459204.8949 & 0.0056 & N \\
150151262 & TOI-712.02 & 23 & 2460135.465 & 0.005 & N \\
150151262 & TOI-712.03 & -2 & 2458470.8605 & 0.0047 & N \\
150151262 & TOI-712.03 & 6 & 2459149.6010 & 0.0037 & N \\
150151262 & TOI-712.03 & 7 & 2459234.3906 & 0.0051 & N \\
150151262 & TOI-712.03 & 16 & 2459997.9444 & 0.0034 & N \\
150151262 & TOI-712.03 & 17 & 2460082.8170 & 0.0042 & N \\
207425167 & TOI-1812.01 & -16 & 2458716.6377 & 0.0074 & Y \\
207425167 & TOI-1812.01 & -14 & 2458739.854 & 0.012 & Y \\
207425167 & TOI-1812.01 & 0 & 2458902.466 & 0.004 & Y \\
207425167 & TOI-1812.01 & 3 & 2458937.3030 & 0.0069 & Y \\
207425167 & TOI-1812.01 & 5 & 2458960.535 & 0.007 & Y \\
207425167 & TOI-1812.01 & 7 & 2458983.7559 & 0.0066 & Y \\
207425167 & TOI-1812.01 & 64 & 2459645.5529 & 0.0048 & Y \\
207425167 & TOI-1812.01 & 69 & 2459703.5694 & 0.0091 & Y \\
207425167 & TOI-1812.01 & 71 & 2459726.7971 & 0.0057 & Y \\
207425167 & TOI-1812.01 & 80 & 2459831.2530 & 0.0046 & Y \\
207425167 & TOI-1812.01 & 85 & 2459889.2677 & 0.0034 & Y \\
207425167 & TOI-1812.02 & -4 & 2458716.1950 & 0.0037 & N \\
207425167 & TOI-1812.02 & 0 & 2458909.5147 & 0.0041 & N \\
207425167 & TOI-1812.02 & 1 & 2458957.8459 & 0.0037 & N \\
207425167 & TOI-1812.02 & 2 & 2459006.1788 & 0.0036 & N \\
207425167 & TOI-1812.02 & 19 & 2459827.78747 & 0.00047 & N \\
219508169 & TOI-2016.01 & -218 & 2458957.619 & 0.003 & N \\
219508169 & TOI-2016.01 & -217 & 2458964.4345 & 0.0072 & N \\
219508169 & TOI-2016.01 & -216 & 2458971.2522 & 0.0021 & N \\
219508169 & TOI-2016.01 & -215 & 2458978.0677 & 0.0029 & N \\
219508169 & TOI-2016.01 & -214 & 2458984.8825 & 0.0051 & N \\
219508169 & TOI-2016.01 & -213 & 2458991.7096 & 0.0079 & N \\
219508169 & TOI-2016.01 & -212 & 2458998.518 & 0.003 & N \\
219508169 & TOI-2016.01 & -211 & 2459005.3431 & 0.0027 & N \\
219508169 & TOI-2016.01 & -109 & 2459700.5916 & 0.0073 & N \\
219508169 & TOI-2016.01 & -107 & 2459714.2052 & 0.0039 & N \\
219508169 & TOI-2016.01 & -105 & 2459727.8471 & 0.0025 & N \\
219508169 & TOI-2016.01 & -104 & 2459734.6560 & 0.0032 & N \\
219508169 & TOI-2016.01 & -103 & 2459741.4718 & 0.0043 & N \\
219508169 & TOI-2016.02 & -618 & 2458957.7462 & 0.0016 & Y \\
219508169 & TOI-2016.02 & -607 & 2458984.7978 & 0.0011 & Y \\
219508169 & TOI-2016.02 & -319 & 2459692.99820 & 0.00099 & Y \\
219508169 & TOI-2016.02 & -308 & 2459720.0471 & 0.0013 & Y \\
219508169 & TOI-2016.03 & -30 & 2458975.3841 & 0.0076 & N \\
219508169 & TOI-2016.03 & -29 & 2459000.333 & 0.004 & N \\
219508169 & TOI-2016.03 & -1 & 2459710.177 & 0.011 & N \\
219508169 & TOI-2016.03 & 0 & 2459735.491 & 0.018 & N \\
251848941 & TOI-178.01 & -280 & 2458360.2397 & 0.0028 & N \\
251848941 & TOI-178.01 & -279 & 2458366.8005 & 0.0019 & N \\
251848941 & TOI-178.01 & -278 & 2458373.3559 & 0.0047 & N \\
251848941 & TOI-178.01 & -277 & 2458379.9152 & 0.0032 & N \\
251848941 & TOI-178.01 & -168 & 2459094.7053 & 0.0019 & N \\
251848941 & TOI-178.01 & -166 & 2459107.8190 & 0.0028 & N \\
251848941 & TOI-178.01 & -2 & 2460183.319 & 0.004 & N \\
251848941 & TOI-178.01 & -1 & 2460189.8848 & 0.0026 & N \\
251848941 & TOI-178.01 & 0 & 2460196.4430 & 0.0029 & N \\
251848941 & TOI-178.01 & 1 & 2460203.0005 & 0.0027 & N \\
260647166 & HD 108236.01 & -52 & 2458571.33466 & 0.00019 & N \\
260647166 & HD 108236.01 & -51 & 2458585.51048 & 0.00023 & N \\
260647166 & HD 108236.01 & 0 & 2459308.48365 & 0.00019 & N \\
260647166 & HD 108236.01 & 1 & 2459322.65707 & 0.00018 & N \\
260647166 & HD 108236.02 & -37 & 2458586.55851 & 0.00022 & N \\
260647166 & HD 108236.02 & -36 & 2458606.15710 & 0.00018 & N \\
260647166 & HD 108236.02 & 0 & 2459311.40278 & 0.00017 & N \\
260647166 & HD 108236.02 & 1 & 2459330.99570 & 0.00017 & N \\
279741379 & TOI-186.01 & -52 & 2458350.3129 & 0.0023 & N \\
279741379 & TOI-186.01 & -32 & 2459062.5828 & 0.0016 & N \\
279741379 & TOI-186.01 & -30 & 2459133.81336 & 0.00029 & N \\
279741379 & TOI-186.01 & 0 & 2460202.21484 & 0.00026 & N \\
279741379 & TOI-186.02 & -240 & 2458331.99 & 0.15 & Y \\
279741379 & TOI-186.02 & -237 & 2458355.6478 & 0.0058 & Y \\
279741379 & TOI-186.02 & -233 & 2458386.8096 & 0.0038 & Y \\
279741379 & TOI-186.02 & -229 & 2458417.9709 & 0.0073 & Y \\
279741379 & TOI-186.02 & -146 & 2459064.514 & 0.034 & Y \\
279741379 & TOI-186.02 & -142 & 2459095.6794 & 0.0061 & Y \\
279741379 & TOI-186.02 & -139 & 2459119.0531 & 0.0036 & Y \\
279741379 & TOI-186.02 & -124 & 2459235.8949 & 0.0064 & Y \\
279741379 & TOI-186.02 & -30 & 2459968.131 & 0.017 & Y \\
279741379 & TOI-186.02 & -6 & 2460155.0850 & 0.0044 & Y \\
279741379 & TOI-186.02 & -2 & 2460186.2564 & 0.0053 & Y \\
288636342 & TOI-1692.02 & -32 & 2458686.8309 & 0.0019 & N \\
288636342 & TOI-1692.02 & -31 & 2458719.031 & 0.002 & N \\
288636342 & TOI-1692.02 & -29 & 2458783.4625 & 0.0021 & N \\
288636342 & TOI-1692.02 & -27 & 2458847.8748 & 0.0019 & N \\
288636342 & TOI-1692.02 & -26 & 2458880.0676 & 0.0019 & N \\
288636342 & TOI-1692.02 & -23 & 2458976.7025 & 0.0018 & N \\
288636342 & TOI-1692.02 & -22 & 2459008.9118 & 0.0019 & N \\
288636342 & TOI-1692.02 & -10 & 2459395.4057 & 0.0019 & N \\
288636342 & TOI-1692.02 & -9 & 2459427.6190 & 0.0021 & N \\
288636342 & TOI-1692.02 & -4 & 2459588.654 & 0.002 & N \\
288636342 & TOI-1692.02 & -3 & 2459620.8617 & 0.0019 & N \\
288636342 & TOI-1692.02 & -1 & 2459685.287 & 0.002 & N \\
288636342 & TOI-1692.02 & 1 & 2459749.6941 & 0.0018 & N \\
288636342 & TOI-1692.02 & 2 & 2459781.9030 & 0.0019 & N \\
288636342 & TOI-1692.02 & 3 & 2459814.1121 & 0.0019 & N \\
288636342 & TOI-1692.02 & 5 & 2459878.5258 & 0.0019 & N \\
288636342 & TOI-1692.02 & 7 & 2459942.9413 & 0.0017 & N \\
308994098 & TOI-790.01 & 0 & 2458352.5011 & 0.0016 & N \\
308994098 & TOI-790.01 & 1 & 2458552.0471 & 0.0024 & N \\
308994098 & TOI-790.01 & 4 & 2459150.7661 & 0.0025 & N \\
308994098 & TOI-790.01 & 5 & 2459350.3952 & 0.0022 & N \\
345143460 & TOI-1533.01 & -457 & 2458780.29752 & 0.00019 & N \\
345143460 & TOI-1533.01 & -456 & 2458783.93493 & 0.00019 & N \\
345143460 & TOI-1533.01 & -455 & 2458787.60565 & 0.00018 & N \\
345143460 & TOI-1533.01 & -408 & 2458958.93349 & 0.00022 & N \\
345143460 & TOI-1533.01 & -407 & 2458962.5794 & 0.0002 & N \\
345143460 & TOI-1533.01 & -405 & 2458969.87099 & 0.00022 & N \\
345143460 & TOI-1533.01 & -404 & 2458973.5193 & 0.0002 & N \\
345143460 & TOI-1533.01 & -403 & 2458977.1716 & 0.0002 & N \\
345143460 & TOI-1533.01 & -402 & 2458980.80834 & 0.00018 & N \\
345143460 & TOI-1533.01 & -162 & 2459855.80143 & 0.00019 & N \\
345143460 & TOI-1533.01 & -161 & 2459859.4473 & 0.0002 & N \\
345143460 & TOI-1533.01 & -160 & 2459863.0931 & 0.0002 & N \\
345143460 & TOI-1533.01 & -159 & 2459866.73888 & 0.00022 & N \\
345143460 & TOI-1533.01 & -158 & 2459870.38471 & 0.00022 & N \\
345143460 & TOI-1533.01 & -157 & 2459874.03047 & 0.00019 & N \\
345143460 & TOI-1533.01 & -156 & 2459877.68800 & 0.00023 & N \\
345143460 & TOI-1533.01 & -155 & 2459881.33651 & 0.00015 & N \\
345143460 & TOI-1533.02 & -73 & 2459853.8879 & 0.0002 & N \\
345143460 & TOI-1533.02 & -72 & 2459861.96995 & 0.00021 & N \\
345143460 & TOI-1533.02 & -71 & 2459870.0335 & 0.0002 & N \\
345143460 & TOI-1533.02 & -70 & 2459878.09699 & 0.00018 & N \\
349972412 & TOI-4504.01 & 0 & 2458483.2002 & 0.0019 & N \\
349972412 & TOI-4504.01 & 1 & 2458565.0996 & 0.0019 & N \\
349972412 & TOI-4504.01 & 2 & 2458647.300 & 0.002 & N \\
349972412 & TOI-4504.01 & 7 & 2459065.2500 & 0.0019 & N \\
349972412 & TOI-4504.01 & 8 & 2459148.500 & 0.002 & N \\
349972412 & TOI-4504.01 & 9 & 2459231.0996 & 0.0021 & N \\
349972412 & TOI-4504.01 & 10 & 2459313.2500 & 0.0018 & N \\
349972412 & TOI-4504.01 & 19 & 2460059.6004 & 0.0021 & N \\
349972412 & TOI-4504.01 & 20 & 2460142.6004 & 0.0018 & N \\
350618622 & TOI-201.01 & -34 & 2458376.05172 & 0.00043 & N \\
350618622 & TOI-201.01 & -33 & 2458429.02976 & 0.00052 & N \\
350618622 & TOI-201.01 & -32 & 2458482.00896 & 0.00048 & N \\
350618622 & TOI-201.01 & -30 & 2458587.96462 & 0.00052 & N \\
350618622 & TOI-201.01 & -21 & 2459064.7688 & 0.0004 & N \\
350618622 & TOI-201.01 & -20 & 2459117.74760 & 0.00038 & N \\
350618622 & TOI-201.01 & -18 & 2459223.70386 & 0.00035 & N \\
350618622 & TOI-201.01 & -17 & 2459276.68107 & 0.00033 & N \\
350618622 & TOI-201.01 & -16 & 2459329.65908 & 0.00038 & N \\
350618622 & TOI-201.01 & -15 & 2459382.63815 & 0.00035 & N \\
350618622 & TOI-201.01 & -4 & 2459965.39390 & 0.00036 & N \\
350618622 & TOI-201.01 & -2 & 2460071.36794 & 0.00037 & N \\
350618622 & TOI-201.01 & 0 & 2460177.33136 & 0.00033 & N \\
415969908 & TOI-233.01 & 0 & 2458365.259 & 0.004 & N \\
415969908 & TOI-233.01 & 1 & 2458376.9316 & 0.0048 & N \\
415969908 & TOI-233.01 & 62 & 2459088.8008 & 0.0026 & N \\
415969908 & TOI-233.01 & 64 & 2459112.1438 & 0.0022 & N \\
415969908 & TOI-233.01 & 93 & 2459450.5756 & 0.0056 & N \\
415969908 & TOI-233.01 & 94 & 2459462.2444 & 0.0069 & N \\
415969908 & TOI-233.01 & 156 & 2460185.7914 & 0.0053 & N \\
415969908 & TOI-233.01 & 157 & 2460197.4522 & 0.0035 & N \\
415969908 & TOI-233.02 & 0 & 2458359.4680 & 0.0032 & Y \\
415969908 & TOI-233.02 & 102 & 2459094.0206 & 0.0014 & Y \\
415969908 & TOI-233.02 & 152 & 2459453.752 & 0.049 & Y \\
415969908 & TOI-233.02 & 254 & 2460188.5540 & 0.0037 & Y \\
\enddata
\end{deluxetable}

\section{Best-Fit Orbital Parameters for All Analyzed TOIs}

\begin{longrotatetable}
\begin{deluxetable}{cccccccc}
\rotate                        
\tablecaption{Light Curve Parameters of Analyzed TOIs\label{tab:fullparams}}

\tablehead{
\colhead{TIC} & \colhead{Planet} & \colhead{$P$ [d]} & \colhead{$T_0$ [BJD]} & \colhead{$R_p/R_{\star}$} & \colhead{$b$} & \colhead{$e$} & \colhead{$\omega$ [$^\circ$]}
}

\startdata
8260536 & TOI-5398.01 & 10.5910$\pm$0.0009 & 2459616.49217 $\pm$ 0.00049 & 0.09071 $\pm$ 0.00098 & 0.41$\pm$0.12 & 0.12$\pm$0.17 & -24.2$\pm$120.0 \\
8260536 & TOI-5398.02 & 4.7729$\pm$0.0011 & 2459628.6191 $\pm$ 0.0011 & 0.03212 $\pm$ 0.00065 & 0.31$\pm$0.19 & 0.17$\pm$0.19 & -67.5$\pm$99.0 \\
20318757 & TOI-1027.01 & 3.2834572$\pm$4.8e-06 & 2460037.8671 $\pm$ 0.0015 & 0.0321 $\pm$ 0.0022 & 0.56$\pm$0.19 & 0.020$\pm$0.025 & -24.8$\pm$130.0 \\
20318757 & TOI-1027.02 & 11.028746$\pm$3.5e-05 & 2460032.5429 $\pm$ 0.0072 & 0.0267 $\pm$ 0.0038 & 0.57$\pm$0.24 & 0.027$\pm$0.029 & -53.7$\pm$120.0 \\
20318757 & TOI-1027.03 & 5.011328$\pm$1.3e-05 & 2460032.5750 $\pm$ 0.0036 & 0.031 $\pm$ 0.005 & 0.7$\pm$0.2 & 0.024$\pm$0.028 & 68.7$\pm$77.0 \\
27491137 & TOI-2076.01 & 10.356359$\pm$2.5e-05 & 2459675.6958 $\pm$ 0.0064 & 0.0269 $\pm$ 0.0058 & 0.41$\pm$0.26 & 0.38$\pm$0.27 & -36.6$\pm$97.0 \\
55652896 & TOI-216.01 & 34.554785$\pm$1.3e-05 & 2458331.477187 $\pm$ 6.8e-05 & 0.125 $\pm$ 0.016 & 0.5$\pm$0.2 & 0.0005$\pm$0.0007 & 65.2$\pm$73.0 \\
55652896 & TOI-216.02 & 17.3890446$\pm$8.7e-06 & 2458320.310523 $\pm$ 5e-05 & 0.098 $\pm$ 0.013 & 0.53$\pm$0.18 & 0.005$\pm$0.004 & -137.7$\pm$77.0 \\
75878355 & TOI-2134.01 & 9.2292080$\pm$6.4e-06 & 2460496.590 $\pm$ 0.002 & 0.0357 $\pm$ 0.0018 & 0.48$\pm$0.21 & 0.26$\pm$0.22 & -17.2$\pm$130.0 \\
79748331 & TOI-1064.01 & 6.443865$\pm$1.8e-05 & 2458656.6674 $\pm$ 0.0032 & 0.0300 $\pm$ 0.0036 & 0.53$\pm$0.22 & 0.29$\pm$0.25 & 53.9$\pm$130.0 \\
79748331 & TOI-1064.02 & 12.226558$\pm$3e-05 & 2458664.4981 $\pm$ 0.0057 & 0.0309 $\pm$ 0.0041 & 0.53$\pm$0.28 & 0.40$\pm$0.28 & 38.5$\pm$120.0 \\
92226327 & TOI-256.02 & 3.7779409$\pm$9.5e-06 & 2459141.105 $\pm$ 0.001 & 0.0499 $\pm$ 0.0034 & 0.22$\pm$0.21 & 0.27$\pm$0.22 & -82.1$\pm$56.0 \\
101011575 & HD 73583.01 & 6.3980629$\pm$6.1e-06 & 2459982.8446 $\pm$ 0.0018 & 0.0378 $\pm$ 0.0018 & 0.5$\pm$0.2 & 0.22$\pm$0.24 & -24.3$\pm$150.0 \\
101011575 & HD 73583.02 & 18.879281$\pm$5.8e-05 & 2459987.3398 $\pm$ 0.0036 & 0.0340 $\pm$ 0.0022 & 0.32$\pm$0.21 & 0.24$\pm$0.24 & -58.5$\pm$67.0 \\
102840239 & TOI-815.01 & 11.197276$\pm$2.4e-05 & 2460026.6223 $\pm$ 0.0022 & 0.0324 $\pm$ 0.0018 & 0.37$\pm$0.21 & 0.23$\pm$0.24 & -61.7$\pm$91.0 \\
130181866 & TOI-1726.01 & 7.1079342$\pm$1.5e-06 & 2458845.374149 $\pm$ 9.7e-05 & 0.0230 $\pm$ 0.0042 & 0.46$\pm$0.22 & 0.000495$\pm$4.9e-05 & 63.3$\pm$2.8 \\
130181866 & TOI-1726.02 & 20.5438161$\pm$5.4e-06 & 2459583.636349 $\pm$ 8e-05 & 0.027 $\pm$ 0.006 & 0.53$\pm$0.19 & 0.000496$\pm$4e-05 & 63.5$\pm$2.6 \\
146413471 & TOI-6454.01 & 22.50129$\pm$0.00016 & 2459948.818 $\pm$ 0.013 & 0.0557 $\pm$ 0.0058 & 0.4$\pm$0.3 & 0.28$\pm$0.24 & -33.4$\pm$140.0 \\
146413471 & TOI-6454.02 & 10.983580$\pm$7.1e-05 & 2459974.386 $\pm$ 0.019 & 0.0376 $\pm$ 0.0073 & 0.36$\pm$0.27 & 0.33$\pm$0.26 & -33.6$\pm$110.0 \\
149601126 & TOI-2525.01 & 23.34181$\pm$0.00017 & 2458310.547579 $\pm$ 9.4e-05 & 0.11 $\pm$ 0.02 & 0.49$\pm$0.19 & 0.000488$\pm$4.4e-05 & 63.1$\pm$2.6 \\
149601126 & TOI-2525.02 & 49.244555$\pm$3.7e-05 & 2458286.13784 $\pm$ 0.00013 & 0.12 $\pm$ 0.02 & 0.52$\pm$0.18 & 0.000494$\pm$4.1e-05 & 63.2$\pm$2.3 \\
153949511 & TOI-1277.02 & 14.855972$\pm$4.7e-05 & 2460319.119 $\pm$ 0.016 & 0.0304 $\pm$ 0.0025 & 0.35$\pm$0.23 & 0.22$\pm$0.25 & -87.8$\pm$70.0 \\
179230828 & TOI-5000.01 & 5.5422295$\pm$2.8e-06 & 2459302.141312 $\pm$ 5.3e-05 & 0.131 $\pm$ 0.025 & 0.54$\pm$0.17 & 0.000499$\pm$3.4e-05 & 63.0$\pm$2.0 \\
179230828 & TOI-5000.02 & 15.339286$\pm$1.4e-05 & 2459285.959514 $\pm$ 8.2e-05 & 0.105 $\pm$ 0.021 & 0.50$\pm$0.17 & 0.000493$\pm$4.7e-05 & 63.4$\pm$2.3 \\
232608943 & TOI-4600.01 & 82.6901$\pm$0.0002 & 2459419.395890 $\pm$ 7.3e-05 & 0.079 $\pm$ 0.016 & 0.51$\pm$0.18 & 0.000496$\pm$5.1e-05 & 63.8$\pm$3.0 \\
254113311 & TOI-1130.01 & 8.3499796$\pm$2e-06 & 2458657.908389 $\pm$ 9.6e-05 & 0.134 $\pm$ 0.027 & 0.90$\pm$0.02 & 0.000497$\pm$4.2e-05 & 63.3$\pm$2.4 \\
254113311 & TOI-1130.02 & 4.068637$\pm$6e-05 & 2458658.691936 $\pm$ 6.6e-05 & 0.0456 $\pm$ 0.0089 & 0.5$\pm$0.2 & 0.000500$\pm$5.2e-05 & 63.6$\pm$2.8 \\
261257684 & TOI-904.01 & 10.877275$\pm$1.1e-05 & 2458626.9679 $\pm$ 0.0018 & 0.0383 $\pm$ 0.0018 & 0.31$\pm$0.18 & 0.16$\pm$0.19 & -51.6$\pm$81.0 \\
261257684 & TOI-904.02 & 83.99935$\pm$0.00024 & 2458630.35 $\pm$ 0.01 & 0.0401 $\pm$ 0.0024 & 0.37$\pm$0.24 & 0.26$\pm$0.24 & -23.9$\pm$120.0 \\
273231214 & TOI-4581.01 & 22.342871$\pm$8.5e-05 & 2458719.3654 $\pm$ 0.0048 & 0.0505 $\pm$ 0.0021 & 0.50$\pm$0.18 & 0.20$\pm$0.23 & -18.3$\pm$140.0 \\
286864983 & TOI-772.01 & 11.0163419$\pm$2.2e-06 & 2458575.94471 $\pm$ 0.00014 & 0.078 $\pm$ 0.017 & 0.49$\pm$0.18 & 0.000492$\pm$3.7e-05 & 63.6$\pm$2.8 \\
286864983 & TOI-772.02 & 744.20081$\pm$0.00056 & 2459313.64135 $\pm$ 0.00018 & 0.061 $\pm$ 0.012 & 0.66$\pm$0.15 & 0.000496$\pm$4.2e-05 & 63.7$\pm$2.7 \\
306472057 & TOI-791.01 & 139.30478$\pm$0.00019 & 2458427.6228 $\pm$ 0.0016 & 0.06440 $\pm$ 0.00071 & 0.19$\pm$0.15 & 0.23$\pm$0.25 & -17.2$\pm$110.0 \\
306996324 & TOI-776.01 & 15.6653323$\pm$7.3e-06 & 2458572.59943 $\pm$ 0.00012 & 0.0349 $\pm$ 0.0068 & 0.5$\pm$0.2 & 0.000490$\pm$4.9e-05 & 63.2$\pm$2.3 \\
306996324 & TOI-776.02 & 8.2466137$\pm$2.9e-06 & 2458571.42749 $\pm$ 0.00011 & 0.0295 $\pm$ 0.0054 & 0.45$\pm$0.21 & 0.000509$\pm$4.6e-05 & 64.2$\pm$2.4 \\
307809773 & TOI-4599.01 & 2.7694943$\pm$3.3e-06 & 2460284.459 $\pm$ 0.027 & 0.023 $\pm$ 0.002 & 0.54$\pm$0.13 & 0.024$\pm$0.041 & 48.5$\pm$97.0 \\
307809773 & TOI-4599.02 & 5.7060867$\pm$5.4e-06 & 2460277.800 $\pm$ 0.025 & 0.0314 $\pm$ 0.0042 & 0.862$\pm$0.057 & 0.047$\pm$0.065 & 3.7$\pm$100.0 \\
347332255 & HD 110067.01 & 9.1136778$\pm$2.1e-06 & 2459640.15796 $\pm$ 0.00013 & 0.0262 $\pm$ 0.0039 & 0.36$\pm$0.21 & 0.000497$\pm$3.8e-05 & 63.4$\pm$2.9 \\
347332255 & HD 110067.02 & 13.6736945$\pm$2.3e-06 & 2459657.45702 $\pm$ 0.00014 & 0.0266 $\pm$ 0.0046 & 0.28$\pm$0.12 & 0.000506$\pm$4.2e-05 & 63.6$\pm$3.0 \\
374180079 & K2-266.01 & 14.698084$\pm$5.1e-05 & 2459267.7 $\pm$ 0.1 & 0.0347 $\pm$ 0.0029 & 0.29$\pm$0.21 & 0.25$\pm$0.24 & -34.2$\pm$90.0 \\
374180079 & K2-266.02 & 19.48351$\pm$0.00018 & 2459263.381 $\pm$ 0.013 & 0.0325 $\pm$ 0.0029 & 0.4$\pm$0.2 & 0.30$\pm$0.24 & -42.9$\pm$98.0 \\
384984325 & TOI-6109.01 & 8.538663$\pm$3.3e-05 & 2459908.0959 $\pm$ 0.0074 & 0.0297 $\pm$ 0.0097 & 0.59$\pm$0.36 & 0.41$\pm$0.26 & 41.8$\pm$95.0 \\
384984325 & TOI-6109.02 & 5.69534$\pm$0.00062 & 2459900.8161 $\pm$ 0.0073 & 0.030 $\pm$ 0.014 & 0.65$\pm$0.29 & 0.44$\pm$0.28 & 0.1$\pm$140.0 \\
440887364 & TOI-836.01 & 8.595475$\pm$3.1e-05 & 2459356.1624 $\pm$ 0.0049 & 0.0292 $\pm$ 0.0044 & 0.46$\pm$0.28 & 0.29$\pm$0.22 & -20.5$\pm$130.0 \\
440887364 & TOI-836.02 & 3.816725$\pm$1.2e-05 & 2459355.7073 $\pm$ 0.0039 & 0.0199 $\pm$ 0.0046 & 0.5$\pm$0.3 & 0.41$\pm$0.28 & 7.3$\pm$140.0 \\
453211454 & HD 63935.01 & 9.058820$\pm$1.2e-05 & 2460269.9740 $\pm$ 0.0032 & 0.0282 $\pm$ 0.0012 & 0.36$\pm$0.22 & 0.24$\pm$0.22 & -54.7$\pm$95.0 \\
27064468 & TOI-5126.01 & 5.458976$\pm$7.3e-05 & 2460282.100 $\pm$ 0.014 & 0.034 $\pm$ 0.001 & 0.37$\pm$0.21 & 0.22$\pm$0.24 & -31.7$\pm$130.0 \\
31374837 & TOI-431.01 & 12.461005$\pm$1.1e-05 & 2458440.6299 $\pm$ 0.0019 & 0.0428 $\pm$ 0.0024 & 0.39$\pm$0.19 & 0.30$\pm$0.27 & -40.2$\pm$98.0 \\
34077285 & TOI-880.01 & 6.3872508$\pm$8.5e-06 & 2459224.8311 $\pm$ 0.0012 & 0.0470 $\pm$ 0.0017 & 0.5$\pm$0.2 & 0.24$\pm$0.18 & 47.0$\pm$78.0 \\
36724087 & LTT 3780.01 & 12.2533$\pm$0.0015 & 2459600.5420 $\pm$ 0.0011 & 0.0399 $\pm$ 0.0038 & 0.77$\pm$0.22 & 0.42$\pm$0.24 & 45.1$\pm$80.0 \\
94986319 & TOI-421.01 & 16.067521$\pm$1.5e-05 & 2459195.31533 $\pm$ 0.00015 & 0.04523 $\pm$ 0.00082 & 0.978$\pm$0.028 & 0.606$\pm$0.086 & -27.3$\pm$11.0 \\
120826158 & TOI-4495.01 & 5.183000$\pm$3.9e-05 & 2459765.92489 $\pm$ 0.00095 & 0.0269 $\pm$ 0.0011 & 0.63$\pm$0.16 & 0.3$\pm$0.2 & 27.9$\pm$120.0 \\
120826158 & TOI-4495.02 & 2.569366$\pm$2.4e-05 & 2459770.0038 $\pm$ 0.0027 & 0.01681 $\pm$ 0.00053 & 0.38$\pm$0.22 & 0.21$\pm$0.22 & -11.4$\pm$150.0 \\
120896927 & HD 15337.01 & 17.180725$\pm$4.3e-05 & 2459118.9568 $\pm$ 0.0015 & 0.0247 $\pm$ 0.0015 & 0.87$\pm$0.23 & 0.53$\pm$0.22 & 85.9$\pm$59.0 \\
127530399 & TOI-822.01 & 7.1335189$\pm$5.8e-06 & 2459359.00854 $\pm$ 0.00015 & 0.11744 $\pm$ 0.00071 & 0.2$\pm$0.1 & 0.14$\pm$0.12 & 28.6$\pm$120.0 \\
142087638 & TOI-2404.01 & 20.36271$\pm$0.00011 & 2459384.851 $\pm$ 0.002 & 0.0293 $\pm$ 0.0018 & 0.554$\pm$0.097 & 0.000501$\pm$4e-05 & 63.1$\pm$2.4 \\
142087638 & TOI-2404.02 & 74.6049$\pm$0.0023 & 2458558.9733 $\pm$ 0.0037 & 0.0678 $\pm$ 0.0059 & 0.925$\pm$0.018 & 0.000497$\pm$5.2e-05 & 63.1$\pm$2.8 \\
144401492 & TOI-1803.01 & 12.885865$\pm$5.2e-05 & 2459659.0520 $\pm$ 0.0018 & 0.0499 $\pm$ 0.0027 & 0.71$\pm$0.24 & 0.42$\pm$0.21 & 73.4$\pm$91.0 \\
150151262 & TOI-712.01 & 9.531366$\pm$1.3e-05 & 2458757.2688 $\pm$ 0.0011 & 0.0326 $\pm$ 0.0037 & 0.901$\pm$0.066 & 0.17$\pm$0.13 & 16.0$\pm$120.0 \\
150151262 & TOI-712.02 & 51.69916$\pm$0.00014 & 2458946.3993 $\pm$ 0.0029 & 0.0239 $\pm$ 0.0038 & 0.62$\pm$0.14 & 0.24$\pm$0.22 & -58.3$\pm$120.0 \\
150151262 & TOI-712.03 & 84.83872$\pm$0.00038 & 2458640.5465 $\pm$ 0.0032 & 0.0398 $\pm$ 0.0013 & 0.18$\pm$0.12 & 0.26$\pm$0.21 & -71.5$\pm$58.0 \\
153065527 & TOI-406.01 & 13.175678$\pm$3e-05 & 2458388.5669 $\pm$ 0.0017 & 0.0370 $\pm$ 0.0019 & 0.47$\pm$0.17 & 0.15$\pm$0.16 & 29.0$\pm$150.0 \\
153065527 & TOI-406.02 & 3.307446$\pm$2.9e-05 & 2458385.3910 $\pm$ 0.0022 & 0.0284 $\pm$ 0.0023 & 0.65$\pm$0.24 & 0.41$\pm$0.19 & 68.4$\pm$100.0 \\
178819686 & TOI-763.01 & 5.6057736$\pm$9.9e-06 & 2460046.4227 $\pm$ 0.0022 & 0.02048 $\pm$ 0.00096 & 0.4$\pm$0.2 & 0.22$\pm$0.25 & -67.6$\pm$99.0 \\
178819686 & TOI-763.02 & 12.277248$\pm$3e-05 & 2460046.2344 $\pm$ 0.0034 & 0.0248 $\pm$ 0.0013 & 0.37$\pm$0.22 & 0.24$\pm$0.24 & -39.0$\pm$100.0 \\
207425167 & TOI-1812.02 & 11.609806$\pm$4.5e-05 & 2458902.4638 $\pm$ 0.0025 & 0.0323 $\pm$ 0.0016 & 0.45$\pm$0.24 & 0.17$\pm$0.18 & 29.8$\pm$140.0 \\
207425167 & TOI-1812.03 & 48.333$\pm$0.004 & 2458909.5161 $\pm$ 0.0018 & 0.0667 $\pm$ 0.0054 & 0.63$\pm$0.23 & 0.62$\pm$0.25 & -52.0$\pm$23.0 \\
207468071 & TOI-1836.01 & 20.38068$\pm$0.00022 & 2460461.7296 $\pm$ 0.0014 & 0.04683 $\pm$ 0.00064 & 0.5$\pm$0.1 & 0.13$\pm$0.14 & -1.0$\pm$140.0 \\
207468071 & TOI-1836.02 & 1.7727509$\pm$4.2e-06 & 2460477.1317 $\pm$ 0.0045 & 0.0146 $\pm$ 0.0012 & 0.85$\pm$0.15 & 0.57$\pm$0.25 & 78.1$\pm$100.0 \\
219508169 & TOI-2016.01 & 6.816123$\pm$2.1e-05 & 2460443.5357 $\pm$ 0.0033 & 0.0331 $\pm$ 0.0014 & 0.4$\pm$0.2 & 0.22$\pm$0.21 & -11.4$\pm$120.0 \\
219508169 & TOI-2016.02 & 2.4590369$\pm$3.8e-06 & 2460477.4290 $\pm$ 0.0019 & 0.02878 $\pm$ 0.00099 & 0.49$\pm$0.18 & 0.2$\pm$0.2 & 7.6$\pm$140.0 \\
219508169 & TOI-2016.03 & 25.33691$\pm$0.00039 & 2459735.4989 $\pm$ 0.0094 & 0.0315 $\pm$ 0.0019 & 0.3$\pm$0.2 & 0.39$\pm$0.21 & -92.1$\pm$43.0 \\
234345288 & TOI-213.01 & 23.519906$\pm$5.8e-05 & 2460177.8868 $\pm$ 0.0015 & 0.0338 $\pm$ 0.0011 & 0.45$\pm$0.21 & 0.22$\pm$0.23 & 5.1$\pm$160.0 \\
234345288 & TOI-213.02 & 7.756554$\pm$3.2e-05 & 2458328.4354 $\pm$ 0.0038 & 0.01749 $\pm$ 0.00078 & 0.34$\pm$0.19 & 0.20$\pm$0.18 & -13.1$\pm$120.0 \\
243185500 & TOI-1468.01 & 15.532431$\pm$3.5e-05 & 2459450.35556 $\pm$ 0.00067 & 0.0488 $\pm$ 0.0015 & 0.65$\pm$0.16 & 0.32$\pm$0.18 & 37.6$\pm$110.0 \\
243185500 & TOI-1468.02 & 1.8805297$\pm$6.8e-06 & 2459448.30618 $\pm$ 0.00044 & 0.0334 $\pm$ 0.0011 & 0.4$\pm$0.2 & 0.2$\pm$0.2 & -23.2$\pm$130.0 \\
251848941 & TOI-178.01 & 6.557854$\pm$1e-05 & 2460196.4375 $\pm$ 0.0014 & 0.0334 $\pm$ 0.0013 & 0.48$\pm$0.22 & 0.25$\pm$0.21 & 16.0$\pm$140.0 \\
257605131 & TOI-451.01 & 16.364860$\pm$4e-05 & 2458416.6356 $\pm$ 0.0018 & 0.031 $\pm$ 0.001 & 0.4$\pm$0.2 & 0.22$\pm$0.23 & -3.7$\pm$150.0 \\
259377017 & TOI-270.01 & 5.6605688$\pm$3.7e-06 & 2459198.96185 $\pm$ 0.00027 & 0.05765 $\pm$ 0.00094 & 0.31$\pm$0.19 & 0.13$\pm$0.14 & 13.8$\pm$120.0 \\
259377017 & TOI-270.02 & 11.379567$\pm$1.5e-05 & 2458389.67974 $\pm$ 0.00059 & 0.0500 $\pm$ 0.0011 & 0.34$\pm$0.17 & 0.12$\pm$0.17 & -19.0$\pm$120.0 \\
259377017 & TOI-270.03 & 3.3601373$\pm$7.6e-06 & 2458383.73248 $\pm$ 0.00078 & 0.03007 $\pm$ 0.00079 & 0.27$\pm$0.19 & 0.16$\pm$0.15 & -28.2$\pm$140.0 \\
260647166 & HD 108236.01 & 14.1758961$\pm$8.4e-06 & 2459308.48235 $\pm$ 0.00013 & 0.0274 $\pm$ 0.0052 & 0.50$\pm$0.17 & 0.000499$\pm$3.9e-05 & 62.9$\pm$2.3 \\
260647166 & HD 108236.02 & 19.5901558$\pm$9.1e-06 & 2459311.40415 $\pm$ 0.00012 & 0.0313 $\pm$ 0.0069 & 0.52$\pm$0.25 & 0.000510$\pm$4.9e-05 & 62.6$\pm$2.3 \\
281837575 & TOI-5143.01 & 5.20925$\pm$0.00031 & 2459553.29121 $\pm$ 0.00044 & 0.112 $\pm$ 0.021 & 0.951$\pm$0.034 & 0.14$\pm$0.11 & 4.2$\pm$150.0 \\
282576340 & TOI-2494.01 & 8.376130$\pm$1.2e-05 & 2459210.73494 $\pm$ 0.00079 & 0.116 $\pm$ 0.014 & 0.897$\pm$0.036 & 0.567$\pm$0.078 & -41.8$\pm$7.0 \\
288636342 & TOI-1692.02 & 32.207967$\pm$1.8e-05 & 2459717.4871 $\pm$ 0.0006 & 0.0470 $\pm$ 0.0094 & 0.49$\pm$0.19 & 0.000501$\pm$3.8e-05 & 63.5$\pm$2.7 \\
308994098 & TOI-790.01 & 199.5778$\pm$0.0009 & 2458352.4917 $\pm$ 0.0014 & 0.03673 $\pm$ 0.00062 & 0.42$\pm$0.23 & 0.32$\pm$0.16 & 73.9$\pm$60.0 \\
318022259 & TOI-1730.01 & 6.226228$\pm$2.6e-05 & 2459952.6296 $\pm$ 0.0016 & 0.0343 $\pm$ 0.0018 & 0.36$\pm$0.21 & 0.20$\pm$0.24 & -22.1$\pm$130.0 \\
345143460 & TOI-1533.01 & 3.6458045$\pm$1.4e-06 & 2460446.42410 $\pm$ 0.00012 & 0.0316 $\pm$ 0.0058 & 0.50$\pm$0.18 & 0.000506$\pm$4.8e-05 & 63.6$\pm$2.2 \\
345143460 & TOI-1533.02 & 8.0637952$\pm$2.2e-06 & 2460442.9243 $\pm$ 0.0062 & 0.098 $\pm$ 0.017 & 0.67$\pm$0.18 & 0.000497$\pm$4.6e-05 & 63.3$\pm$2.9 \\
349972412 & TOI-4504.01 & 82.92940$\pm$0.00063 & 2458482.984 $\pm$ 0.001 & 0.110 $\pm$ 0.024 & 0.52$\pm$0.18 & 0.000494$\pm$4.7e-05 & 63.1$\pm$2.7 \\
350293646 & WASP-84.01 & 8.5234957$\pm$2.6e-06 & 2459985.30409 $\pm$ 0.00029 & 0.127 $\pm$ 0.001 & 0.727$\pm$0.058 & 0.136$\pm$0.091 & 43.1$\pm$110.0 \\
350618622 & TOI-201.01 & 52.978142$\pm$1.6e-05 & 2460177.32060 $\pm$ 0.00021 & 0.07884 $\pm$ 0.00057 & 0.746$\pm$0.012 & 0.010$\pm$0.012 & 37.8$\pm$37.0 \\
352239069 & TOI-1404.01 & 6.867409$\pm$3.3e-05 & 2459908.5092 $\pm$ 0.0025 & 0.02093 $\pm$ 0.00076 & 0.26$\pm$0.18 & 0.2$\pm$0.2 & -48.5$\pm$78.0 \\
352239069 & TOI-1404.02 & 14.431886$\pm$4.7e-05 & 2459908.5231 $\pm$ 0.0012 & 0.0402 $\pm$ 0.0015 & 0.59$\pm$0.12 & 0.19$\pm$0.18 & -7.0$\pm$130.0 \\
352682207 & TOI-4010.01 & 5.414621$\pm$6.1e-05 & 2459720.20042 $\pm$ 0.00068 & 0.0606 $\pm$ 0.0021 & 0.5$\pm$0.2 & 0.20$\pm$0.15 & 6.9$\pm$130.0 \\
352682207 & TOI-4010.02 & 14.70914$\pm$0.00023 & 2459721.4262 $\pm$ 0.0013 & 0.0603 $\pm$ 0.0027 & 0.68$\pm$0.12 & 0.25$\pm$0.22 & -1.0$\pm$150.0 \\
352682207 & TOI-4010.03 & 1.348305$\pm$2.2e-05 & 2459741.0398 $\pm$ 0.0018 & 0.0302 $\pm$ 0.0017 & 0.62$\pm$0.22 & 0.35$\pm$0.24 & 32.9$\pm$140.0 \\
355867695 & TOI-1260.01 & 3.1274647$\pm$7.2e-06 & 2459634.7620 $\pm$ 0.0014 & 0.027 $\pm$ 0.001 & 0.43$\pm$0.19 & 0.21$\pm$0.21 & -7.2$\pm$130.0 \\
355867695 & TOI-1260.02 & 7.493162$\pm$1.5e-05 & 2458686.1187 $\pm$ 0.0016 & 0.0355 $\pm$ 0.0017 & 0.72$\pm$0.21 & 0.28$\pm$0.19 & 31.6$\pm$87.0 \\
356158613 & TOI-1449.02 & 2.3691867$\pm$9.7e-06 & 2458683.5846 $\pm$ 0.0045 & 0.019 $\pm$ 0.002 & 0.45$\pm$0.31 & 0.2$\pm$0.2 & -10.8$\pm$120.0 \\
356867115 & TOI-1301.01 & 6.0964077$\pm$6.4e-06 & 2459710.29443 $\pm$ 0.00063 & 0.0267 $\pm$ 0.0011 & 0.53$\pm$0.23 & 0.26$\pm$0.18 & 48.5$\pm$130.0 \\
360630575 & HD 109833.01 & 9.188503$\pm$2.5e-05 & 2459376.4111 $\pm$ 0.0011 & 0.01931 $\pm$ 0.00085 & 0.65$\pm$0.15 & 0.32$\pm$0.21 & 18.5$\pm$120.0 \\
360630575 & HD 109833.02 & 13.9028$\pm$0.0014 & 2459344.4641 $\pm$ 0.0014 & 0.0227 $\pm$ 0.0017 & 0.86$\pm$0.18 & 0.58$\pm$0.25 & 78.9$\pm$110.0 \\
368435330 & TOI-1797.01 & 3.645159$\pm$1.4e-05 & 2459631.90206 $\pm$ 0.00098 & 0.02193 $\pm$ 0.00081 & 0.54$\pm$0.25 & 0.36$\pm$0.21 & 40.2$\pm$88.0 \\
371188886 & TOI-2000.01 & 9.1270426$\pm$7.3e-06 & 2460031.89704 $\pm$ 0.00057 & 0.0541 $\pm$ 0.0013 & 0.64$\pm$0.16 & 0.27$\pm$0.19 & 55.8$\pm$130.0 \\
371188886 & TOI-2000.02 & 3.098365$\pm$1.3e-05 & 2460038.8233 $\pm$ 0.0028 & 0.017 $\pm$ 0.001 & 0.70$\pm$0.26 & 0.42$\pm$0.16 & 72.7$\pm$71.0 \\
377064495 & TOI-561.01 & 10.778833$\pm$3.1e-05 & 2460273.2258 $\pm$ 0.0047 & 0.0293 $\pm$ 0.0015 & 0.30$\pm$0.19 & 0.26$\pm$0.26 & -90.4$\pm$71.0 \\
377064495 & TOI-561.02 & 0.44657129$\pm$6e-07 & 2460285.4659 $\pm$ 0.0027 & 0.01457 $\pm$ 0.00055 & 0.27$\pm$0.19 & 0.29$\pm$0.25 & -114.9$\pm$69.0 \\
415969908 & TOI-233.01 & 11.670028$\pm$1.4e-05 & 2458365.2606 $\pm$ 0.0023 & 0.0478 $\pm$ 0.0027 & 0.36$\pm$0.23 & 0.16$\pm$0.16 & -12.3$\pm$110.0 \\
415969908 & TOI-233.02 & 7.201137$\pm$1.4e-05 & 2458359.4973 $\pm$ 0.0024 & 0.0472 $\pm$ 0.0026 & 0.5$\pm$0.2 & 0.23$\pm$0.22 & -41.4$\pm$130.0 \\
425997655 & HD 23472.01 & 17.6671682$\pm$6.4e-06 & 2460180.3776 $\pm$ 0.0011 & 0.0246 $\pm$ 0.0046 & 0.49$\pm$0.17 & 0.000506$\pm$4e-05 & 63.5$\pm$2.5 \\
425997655 & HD 23472.02 & 29.797520$\pm$1.8e-05 & 2460157.9518 $\pm$ 0.0013 & 0.0216 $\pm$ 0.0041 & 0.48$\pm$0.19 & 0.000504$\pm$4.6e-05 & 63.2$\pm$2.8 \\
425997655 & HD 23472.03 & 12.1623323$\pm$9.4e-06 & 2460184.414 $\pm$ 0.001 & 0.0154 $\pm$ 0.0031 & 0.5$\pm$0.2 & 0.000489$\pm$4.6e-05 & 63.5$\pm$2.3 \\
425997655 & HD 23472.04 & 3.9766837$\pm$2.3e-06 & 2460203.0179 $\pm$ 0.0011 & 0.0101 $\pm$ 0.0019 & 0.51$\pm$0.18 & 0.000501$\pm$4.4e-05 & 63.4$\pm$2.6 \\
425997655 & HD 23472.05 & 7.9076225$\pm$5.3e-06 & 2460197.0386 $\pm$ 0.0012 & 0.0129 $\pm$ 0.0025 & 0.48$\pm$0.19 & 0.000496$\pm$4.8e-05 & 63.3$\pm$2.3 \\
441739020 & TOI-1670.01 & 40.750139$\pm$1.7e-05 & 2460503.1392 $\pm$ 0.0006 & 0.07555 $\pm$ 0.00091 & 0.776$\pm$0.033 & 0.052$\pm$0.046 & 8.5$\pm$140.0 \\
441739020 & TOI-1670.02 & 10.983690$\pm$4.8e-05 & 2460468.3729 $\pm$ 0.0052 & 0.01567 $\pm$ 0.00095 & 0.86$\pm$0.09 & 0.31$\pm$0.24 & 55.1$\pm$110.0 \\
441763252 & TOI-4468.01 & 2.7708595$\pm$1.3e-06 & 2459848.55419 $\pm$ 0.00016 & 0.1259 $\pm$ 0.0011 & 0.141$\pm$0.095 & 0.08$\pm$0.08 & -34.6$\pm$140.0 \\
467179528 & TOI-1266.01 & 10.894841$\pm$1.4e-05 & 2459649.75343 $\pm$ 0.00062 & 0.0536 $\pm$ 0.0014 & 0.49$\pm$0.17 & 0.20$\pm$0.18 & 17.4$\pm$150.0 \\
467179528 & TOI-1266.02 & 18.801596$\pm$5.8e-05 & 2459630.0449 $\pm$ 0.0014 & 0.040 $\pm$ 0.002 & 0.69$\pm$0.18 & 0.36$\pm$0.23 & 50.9$\pm$110.0 \\
468979441 & TOI-5493.01 & 24.43834$\pm$0.00013 & 2459235.6098 $\pm$ 0.0014 & 0.0475 $\pm$ 0.0011 & 0.31$\pm$0.21 & 0.19$\pm$0.21 & -47.6$\pm$87.0 \\
468979441 & TOI-5493.02 & 9.404368$\pm$8.9e-05 & 2459230.462 $\pm$ 0.005 & 0.01961 $\pm$ 0.00067 & 0.25$\pm$0.19 & 0.31$\pm$0.23 & -91.6$\pm$62.0 \\
468979441 & TOI-5493.03 & 14.56430$\pm$0.00017 & 2459238.3021 $\pm$ 0.0085 & 0.0184 $\pm$ 0.0018 & 0.36$\pm$0.23 & 0.24$\pm$0.26 & -36.3$\pm$120.0 \\
\enddata

\end{deluxetable}
\end{longrotatetable}

\end{document}